\documentclass[aps,prl,twocolumn,superscriptaddress,longbibliography]{revtex4-1}
\usepackage{mathrsfs}
\usepackage{tabularx}
\usepackage{slashed}
\usepackage{amsmath}
\usepackage{ulem}
\usepackage{amsfonts}
\usepackage{graphicx,color}
\usepackage{times}
\usepackage{braket}
\usepackage[caption=false]{subfig}
\usepackage[colorlinks,linkcolor=red,citecolor=blue]{hyperref}
\newcommand{\comment}[1]{}

\usepackage{amsmath}
\usepackage{amsmath}
\usepackage{relsize}

\usepackage{enumerate}
\usepackage{graphicx}
\usepackage{float}
\usepackage{braket}
\usepackage{dcolumn}
\usepackage{bm}
\usepackage{hyperref}
\hypersetup{colorlinks,linkcolor={blue},citecolor={blue},urlcolor={blue}}  
\usepackage{xcolor} 

\usepackage{fixltx2e}
\AtBeginDocument{%
	\newwrite\bibnotes
	\def\bibnotesext{Notes.bib}
	\immediate\openout\bibnotes=\jobname\bibnotesext
	\immediate\write\bibnotes{@CONTROL{REVTEX41Control}}
	\immediate\write\bibnotes{@CONTROL{%
			apsrev41Control,author="08",editor="1",pages="1",title="1",year="1"}}
	\if@filesw
	\immediate\write\@auxout{\string\citation{apsrev41Control}}%
	\fi
}%

\begin{document}



\title{How quantum selection rules influence the magneto-optical effects of driven, ultrafast magnetization dynamics}

\author{Mohamed F. Elhanoty}
\email{mohamed.elhanoty@physics.uu.se}
 \affiliation{Division of Materials Theory, Department of Physics and Astronomy, Uppsala University, Box-516, SE 75120, Sweden}

\author{Olle Eriksson}
\affiliation{Division of Materials Theory, Department of Physics and Astronomy, Uppsala University, Box-516, SE 75120, Sweden}
\affiliation{Wallenberg Initiative Materials Science for Sustainability (WISE), Uppsala University, 75121 Uppsala, Sweden }

\author{Chin Shen Ong}
 \affiliation{Division of Materials Theory, Department of Physics and Astronomy, Uppsala University, Box-516, SE 75120, Sweden}

\author{Oscar Gr\aa n\"as}
\email{Corresponding Author: oscar.granas@physics.uu.se}
\affiliation{Division of Materials Theory, Department of Physics and Astronomy, Uppsala University, Box-516, SE 75120, Sweden}

\date{\today}
\begin{abstract}
  Ultrafast magnetization dynamics driven by ultrashort pump lasers is typically explained by changes in electronic populations and scattering pathways of excited conduction electrons. This conventional approach overlooks the fundamental role of quantum mechanical selection rules, governing transitions from core states to the conduction band, that forms the key method of the probing step in these experiments. By employing fully \textit{ab initio} time-dependent density functional theory, we reveal that these selection rules profoundly influence the interpretation of ultrafast spin dynamics at specific probe energies. Our analysis for hcp Co and fcc Ni at the M edge demonstrates that the transient dynamics, as revealed in pump-probe experiments, arise from a complex interplay of optical excitations of the M shell. Taking into account the selection rules and conduction electron spin flips, this leads to highly energy-dependent dynamics. These findings address longstanding discrepancies in experimental TMOKE measurements and show that only through meticulous consideration of matrix elements at the probe stage, can one ensure that magnetization dynamics is revealed in its true nature, instead of being muddled by artifacts arising from the choice of probe energy.

\end{abstract}
\maketitle
As we advance toward the next generation of ultrafast and highly efficient devices, the challenge of developing magnetic data storage solutions that can meet the growing demands for speed and reliability becomes increasingly critical~\cite{bhushan2023current,dieny2020opportunities,hirohata2020review}. Leveraging ultrafast magnetization dynamics through the use of ultrashort laser pulses offers a promising approach to revolutionize magnetic data storage. However, uncovering the ultrafast response of magnetic materials to femtosecond laser pulses presents a complex challenge due to the involvement of multiple degrees of freedom, each interacting on different timescales~\cite{shirozhan2024high,kirilyuk2010ultrafast,wang2020ultrafast}.

The pump laser couples directly to the valence and conduction electrons of the material, leading to demagnetization~\cite{kimel20222022,kirilyuk2010ultrafast}. In Ref.~\cite{ryan2023optically}, three key microscopic processes occurring within the first 100 fs are experimentally and theoretically identified: spin flips mediated by spin-orbit coupling (SOC), intrasite spin transfer within the same species, and intersite spin transfer in multicomponent magnetic alloys. Probing these ultrafast processes at such short timescales is highly challenging and requires measurement of the dynamics using probe energies resonant with the excitation of core electrons to unoccupied states in the conduction band~\cite{buades2021attosecond}.

\begin{figure}
    \centering
    \includegraphics[width=\linewidth]{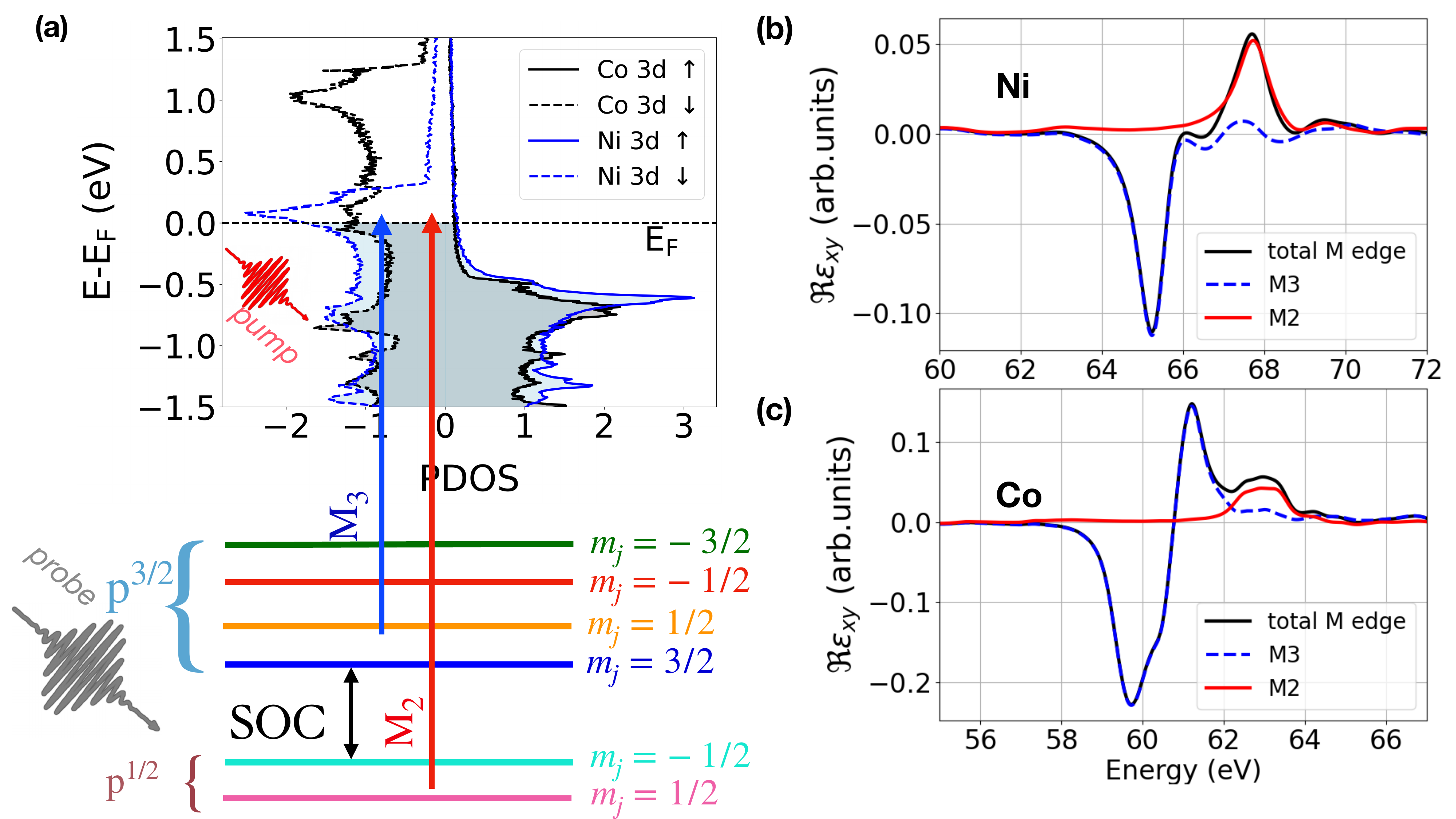}
    \caption{
(a) Schematic diagram illustrating the pump and probe optical excitations and the split of the core states. The upper panel displays the projected density of states (PDOS) for Ni and Co, where the shaded regions below the Fermi (E$_\mathbf{F}$) level represents the ground-state static occupation, and the unshaded region shows the empty states above the Fermi energy.  In the lower panel, the splitting of the 3$p$ core states into $3p^{3/2}$ and $3p^{1/2}$ sublevels, due to spin-orbit coupling (SOC), is shown, with the associated m$_\mathrm{j}$ quantum numbers. Note that the energy scale is not the same for the valence levels and the core levels. The pump laser excites conduction electrons from below to above E$_\mathbf{F}$ and  the probe laser excites core  electrons from the 3$p$ core states to unoccupied states above the Fermi level, E$_{\mathrm{F}}$.
(b,c) The real part of the off-diagonal component of the dielectric tensor in the ground state, depicting the total M edge along with contributions from the $3p^{3/2}$  (M$_3$) and $3p^{1/2}$  (M$_2$) states for Ni (top right panel) and Co (bottom right panel). The calculated $\Re \epsilon_{xy}$ is in a good agreement with photo absorption
cross sections and atomic scattering factor for the experimental measurement in Ref. \cite{henke1993x}. }
\label{fig:schematic}
\end{figure}

Time-resolved pump-probe experiments, such as  time-resolved transverse magneto-optical Kerr effect (TR-TMOKE), have become essential tools for exploring ultrafast magnetization dynamics in both simple elements and magnetic alloys, offering attosecond and femtosecond time resolution~\cite{la2009ultrafast,vodungbo2012laser,tengdin2020direct,ryan2023optically,buades2021attosecond}. For 3$d$ transition metals, TMOKE measurements typically use photon energies resonant with the 3$p$ to 3$d$ transitions at the M-edge~\cite{kirilyuk2010ultrafast}. These energies typically fall within the extreme ultraviolet (XUV) to soft x-ray regimes, coupling the core states to the valence and conduction states. The SOC splits the 3$p$ core levels into $3p^{3/2}$  and $3p^{1/2}$  states ~\cite{de2008core}, resulting in distinct M$_3$ and M$_2$ absorption edges, see Fig.~\ref{fig:schematic}. The calculated real part  of the dielectric tensor, $\Re \epsilon_{xy}$, in Fig. \ref{fig:schematic} agrees well with photo absorption cross sections and atomic scattering factor experiments~\cite{henke1993x}. The energy difference between these levels, typically 1–3~eV in transition metals, leads to a partial overlap of the M$_3$ and M$_2$ edges as shown in  Fig.~\ref{fig:schematic} (b and c). This overlap mixes their respective negative and positive contributions, complicating the interpretation of experimental signals and potentially leading to conflicting interpretation of the measurements. For instance, the XMCD sum rules can not be applied directly to extract spin and orbital moments. This issue is central to the analysis presented in this work.

Recent investigations into the instantaneous magnetization of different elements across varying probe energies revealed energy-dependent ultrafast responses that ostensibly contradicted with one another when different probe energies were used~\cite{hennes2020time,chang2021electron,PhysRevResearch.6.013107,ryan2023optically}. Notably, three studies~\cite{hennes2020time,PhysRevResearch.6.013107,chang2021electron} on Ni have reported starkly different responses within the first 100 fs upon excitation by the same pump pulse. At certain probe energies, reductions in the  transient signal were measured, while at other probe energies, enhancements in the signal were measured. Furthermore, a study on the Heusler compound Co$_2$MnGa~\cite{ryan2023optically} observed a general enhancement of magnetic asymmetry across the entire M edge of Co, serving as an experimental signature of optical intersite spin transfer (OISTR). However, the signal's rates and intensities varied significantly with probe energy.

These findings raise pivotal questions about the interpretations of the enhancements of the magnetic asymmetry measured for simple elements \cite{hennes2020time,chang2021electron,PhysRevResearch.6.013107}, where OISTR simply can not exist, compared to similar effects in magnetic alloys previously attributed to OISTR in theory and experiment~\cite{hofherr2020ultrafast,moller2024verification}. Therefore,
we investigate in this study  whether interpreting TMOKE response changes solely as functions of change in magnetization is valid without being specific about the probe energy. Another fundamental question we address is whether microscopic processes induced by a pump laser, such as optical excitation and spin flips, can manifest unique signatures independently of the probe energy.

In this work, we address these questions through a detailed analysis of the dielectric tensor of face centered cubic (fcc) Ni and hexagonal close packed (hcp) Co at the M-edge in the equilibrium and non-equilibrium, using \textit{ab initio} time-dependent density functional theory (TDDFT). We find that the dynamics induced by the pump laser, including optical excitation and spin flips,  manifest distinct signatures in the transient dielectric tensor and vary with the measuring energy. To identify the microscopic origins of these energy-dependent dynamics, we disentangle the contributions of each core state within the $3p^{3/2}$  and $3p^{1/2}$ manifolds to the total static and time dependent signals of the M$_3$ and M$_2$ edges, respectively. We find that the constituent contributions to the M$_3$ and M$_2$ edges have different optical response to the induced dynamics by the pump laser and their superposition leads to highly energy dependent dynamics in the optical response. Our findings indicate that the observed variations in energy-dependent dynamics, reported in recent ultrafast experiments~\cite{hennes2020time,chang2021electron,PhysRevResearch.6.013107,ryan2023optically}, arise from both the modifications of the electronic structure induced by the pump pulse, as well as details of the probe. The former is found to be influenced by optical excitations that involve occupied and unoccupied valence electron states combined with spin-orbit induced conduction electron spin flips. The latter is demonstrated here to be
the result of a complex interplay of dipole selection rules for the core level to conduction level excitation (primarily $p$ to $d$) that depend on the angular momentum of the core level and the availability of unoccupied conduction electron states combined with the spin character of the states participating in the optical transition. 

The Kerr effect is intrinsically related to the dielectric tensor, $\epsilon$, of a material \cite{oppeneer2001magneto}. Specifically, $M$ is related to $\epsilon_{xy}$ according to the following equation~\cite{oppeneer2001magneto,richter2024relationship}: 
\begin{equation}
    M \propto \Re \bigg[\frac{\epsilon_{xy} \sin{2\theta_i}}{n^4 \cos^2{\theta_i} - n^2 + \sin^2{\theta_i}} \bigg], \label{eq:M_vs_epsilon}
\end{equation}
where $\theta_i$ represents the incident angle of the probe laser, making the calculation of $M$ dependent on the experimental geometry through the Fresnel equations, as analyzed in detail in Ref.~\cite{PhysRevResearch.6.013107}. To simplify our discussion of the magneto-optical effect, we focus on the off-diagonal component of the dielectric tensor, $\epsilon_{xy}$, which is independent of the angle of incidence or transmission (see the methods section in the SM for more details). We excited fcc Ni and hcp Co from their equilibrium states, as shown in Fig.~\ref{fig:schematic}  (see the  details of the pump laser parameters   in the numerical detalis section of the  SM). The pump pulse induced changes in the magnetic spin moment due to spin flips mediated by the SOC such that for Co, it decreased from 1.6~$\mu_{\mathrm{B}}$ per atom to 1.5 $\mu_{\mathrm{B}}$ per atom at 35 fs. For Ni, the magnetic spin moment decreased from 0.61 $\mu_{\mathrm{B}}$ per atom to 0.58 $\mu_{\mathrm{B}}$ per atom.


We conducted our investigation in two stages, by first calculating the linear response relation of  $\Re \epsilon_{xy}$ of  the equilibrium state followed by that in the nonequilibrium state using TDDFT in the adiabatic local spin-density approximation as implemented in the  the full potential ELK code~\cite{elk}.  For each case, we analyzed how individual contributions from each constituent $m_j$ projection of the core states from the $3p^{3/2}$ and $3p^{1/2}$ manifolds contribute to the total intensity of the M$_3$ and M$_2$ edges. This detailed breakdown aims to provide critical insights into how individual core states drive the complex energy-dependent behavior.

 \begin{figure}
    \centering
    \includegraphics[width=\linewidth]{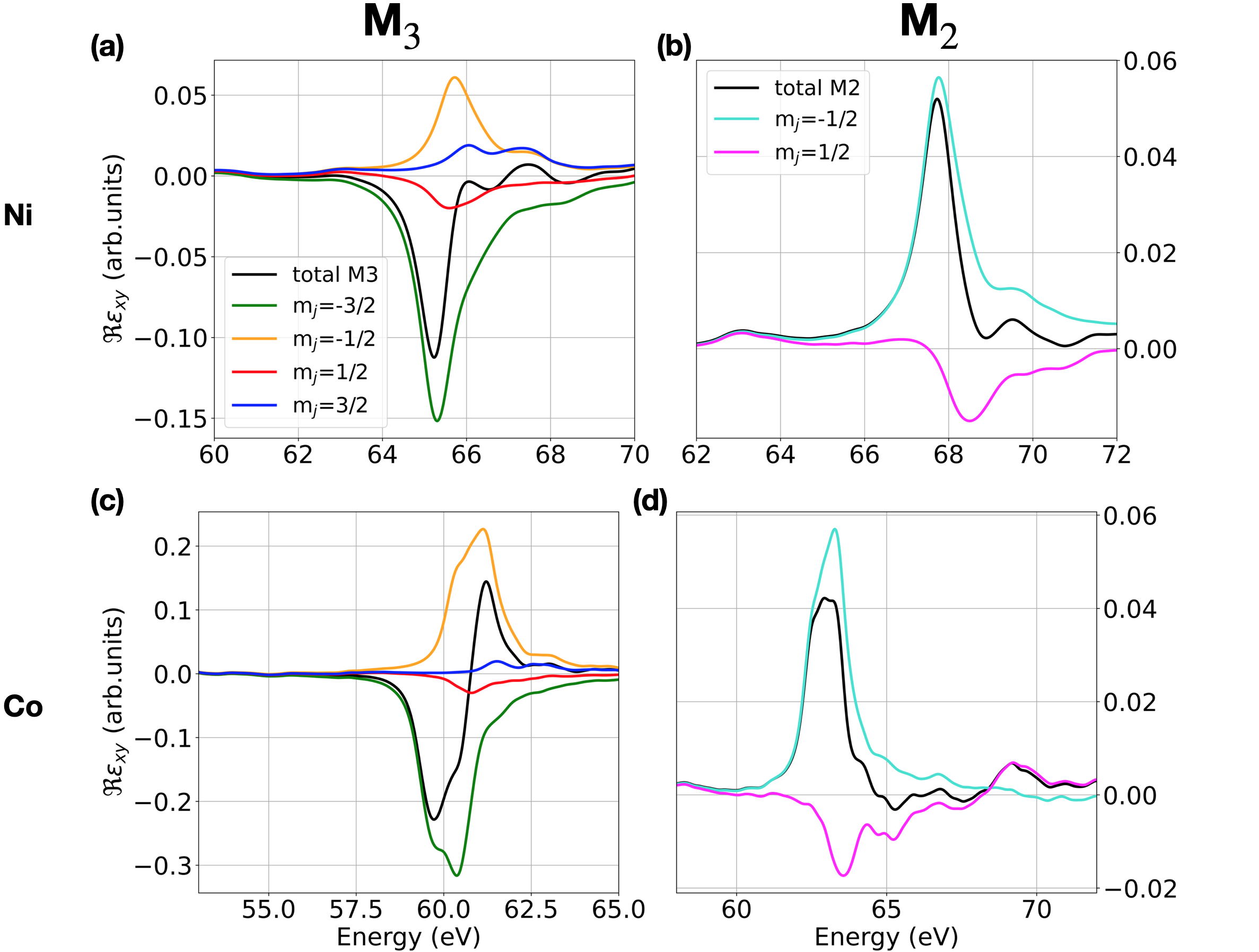}
    \caption{Decomposition of the total M$_3$ and M$_2$ edges for Ni, shown in panels (a) and (b), and Co, shown in panels (c) and (d) into their constituent $3p^{3/2}$  and $3p^{1/2}$  components. For Ni, the $3p^{3/2}$  states are represented by m$_{j=\frac{3}{2}} = {-\frac{3}{2}, -\frac{1}{2}, \frac{1}{2}, \frac{3}{2}}$, and the $3p^{1/2}$  states are represented by m$_{j=\frac{1}{2}} = {-\frac{1}{2}, \frac{1}{2}}$. For Co, the $3p^{3/2}$  and $3p^{1/2}$  components are similarly identified in panels (c) and (d). The colors for the states follow the same list in schematic of Fig.~\ref{fig:schematic}. }
    \label{fig:GSR12stateresolved}
\end{figure}

The M-edge of Ni and Co arises from electronic transitions between core (3$p$) and conduction (3$d$) states, see Fig. \ref{fig:schematic}. Core electrons experience stronger Coulomb attraction and are more localized close to the nucleus of the atom. Thus, they are more pronounced to relativistic effects, particularly spin-orbit coupling (SOC), which couples their orbital ($L$) and spin ($S$) angular momenta. This results in observable spin-orbit splitting between M$_3$ (3$p_{3/2}$) and M$_2$ (3$p_{1/2}$) edges. In contrast, valence electrons, influenced by the crystal field, have weaker SOC, quenching $L$ and making $J$ an unsuitable quantum number, though $S$ remains conserved.

When a probe pulse excites electrons from core to unoccupied conduction states, dipole selection rules dictate these transitions: $\Delta L = \pm 1$, $\Delta S = 0$, and $\Delta J = 0, \pm 1$ (with no transitions between $J=0$ states). As transition intensities depend on initial-final state overlaps, contributions at the M-edge vary by $m_j$, the magnetic quantum number. Clebsch-Gordan coefficients determine how $L$, $S$, and $J$ combine in these transitions, influencing the spectral intensity \cite{edmonds1996angular}.

With this in mind, we now proceed with a detailed examination of the M$_3$ edge. Since the excitation originates from the $3p^{3/2}$ manifold, its intensity is a superposition of contributions from the states of $m_{j=\frac{3}{2}} \in \left\{ -\frac{3}{2}, -\frac{1}{2}, \frac{1}{2}, \frac{3}{2} \right\}$. Figure~\ref{fig:GSR12stateresolved} (a and c) plots the contribution from each $m_j$ state to the total spectral intensities (as shown in Fig.~\ref{fig:schematic}(b and c). Interestingly, contributions from the states with $m_j = \frac{1}{2}$ and $m_j = \frac{3}{2}$ are relatively small and nearly cancel each other out. By decomposing $| j, m_j \rangle$ states using the Clebsch-Gordan coefficients into their $| l, m_l \rangle \otimes | s, m_s \rangle$ components,  we know that both the $m_j = \frac{3}{2}$ and $m_j=\frac{1}{2}$ states are dominated by $m_s = \frac{1}{2}$ components. (In this work, we define $m_s=-\frac{1}{2}$ and $m_s=\frac{1}{2}$ states as spin-down and spin-up states, respectively.) Using the (spin-preserving) selection rule of $\Delta m_s = 0$, we further deduce that the conduction states to which the core states are excited have to possess a dominating $m_s = -\frac{1}{2}$ character. This is further confirmed by our calculations (see Fig.~\ref{fig:schematic}a.) As a result, for the purposes of our discussion, these states will be disregarded from now on. 

Of the remaining $m_j$-states, we note that the magnitude of the $m_j=-\frac{3}{2}$ contribution to the overall M$_3$ peak is significantly larger than that of the $m_j=-\frac{1}{2}$ contribution. This is because the $m_j=-\frac{3}{2}$ state is a pure $m_s=-\frac{1}{2}$ state, whereas the $m_j=-\frac{1}{2}$ state is a linear combination of $m_s = -\frac{1}{2}$ and $m_s = \frac{1}{2}$ states. (see the relevant Clebsch-Gordan coefficients and Fig. \textcolor{blue}{S1} in the SM for the $m_s=\pm\frac{1}{2}$ projection on the core states). Interestingly, even though the M$_3$ peak still has an overall negative sign and that the $m_j=-\frac{3}{2}$ state contributes a negative component, the $m_j=-\frac{1}{2}$ state  contributes a positive component albeit of a smaller magnitude, such that the M$_3$ edge has an overall negative sign.

Similarly, at the M$_2$ edge (see Fig.~\ref{fig:GSR12stateresolved} (b and d)),  the magnitude of the $m_j=-\frac{1}{2}$ contribution is significantly larger than the $m_j=\frac{1}{2}$ contribution because the former has a larger $m_s=-\frac{1}{2}$ component. Even though the M$_2$ edge has an overall positive sign and that the $m_j=-\frac{1}{2}$ state contributes to a positive component, the $m_j=\frac{1}{2}$ state provides a smaller negative contribution, leading to an overall positive signal at the M$_2$ edge.

Summarizing, a few points can be made. First, even though the  $M_3$ ($M_2$) peak has a dominant negative (positive) sign, its constituent $m_j$-projected contributions do not all have the same sign. Second, in terms of magnitude, not all of the $m_j$-projected states contribute equally as well. Most notably, in the region between the M$_3$ and M$_2$ peaks, there are non-trivial overlapping contributions from all of the core $m_j$-states from both the $3p^{3/2}$ and $3p^{1/2}$  manifolds (see Fig. \ref{fig:schematic} (b and c)). This means that when probing the change in TMOKE response, it is imperative that the probe energy be specified since the response is a nontrivial function of the probe energy. As we shall see below, in the nonequilibrium state, time-dependent fluctuations in the signal in this overlapping regime changes dramatically in a way that does not directly relate with the overall decrease in magnetization. For Ni, this overlapping regime ranges from 65 - 68 eV~while for Co, it ranges from 60 - 63~eV.

\begin{figure}
    \centering
    \includegraphics[width=\linewidth]{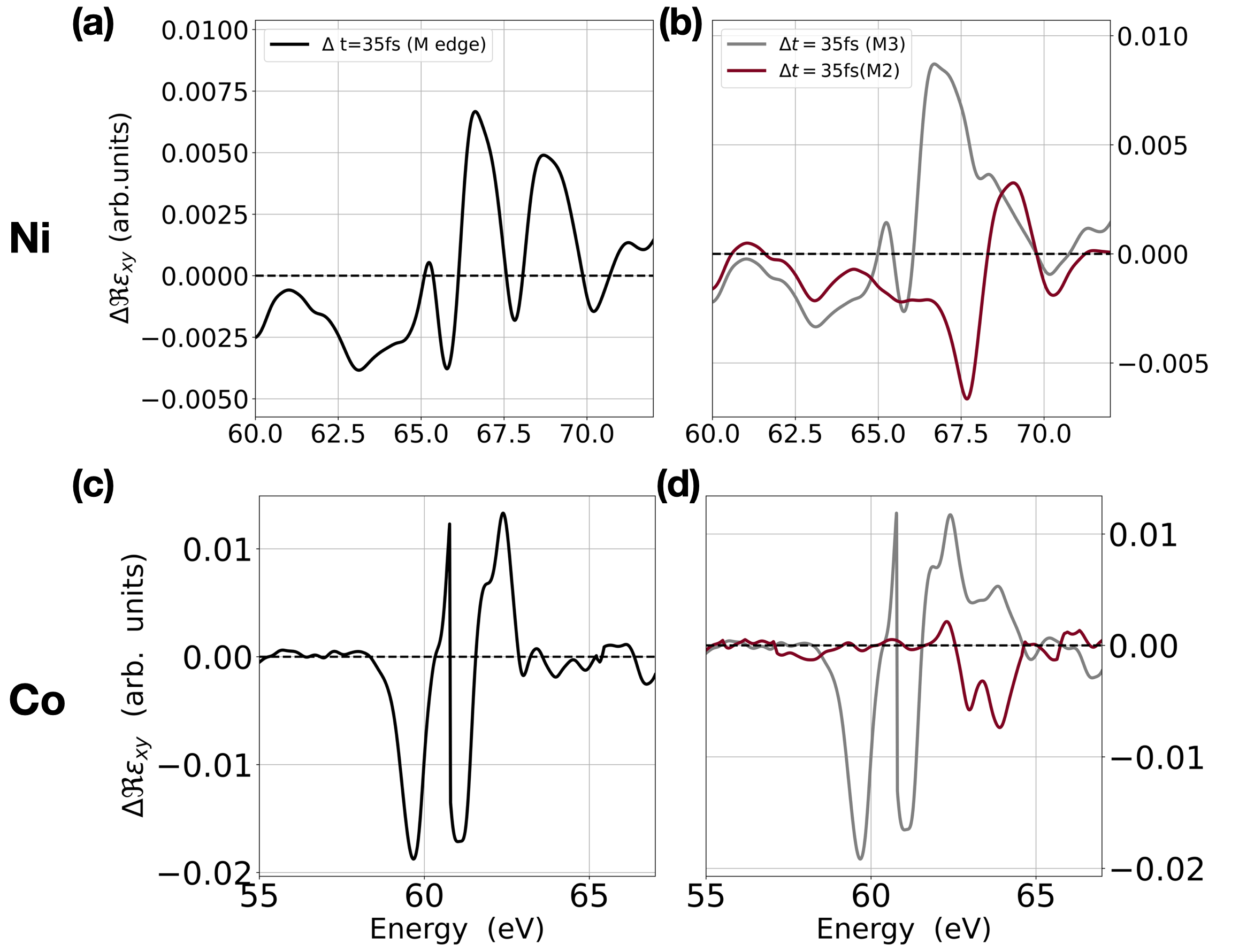}
    \caption{
(a, c) The changes in the real part of the off-diagonal component of the dielectric tensor at 35 fs relative to the ground state for the total M-edge for Ni in (a) and Co in (c).
(b, d) The M-decomposed changes in the $3p^{3/2}$ (M$_3$) and $3p^{1/2}$ (M$_2$) states for Ni in (b) and Co in (d). }

    \label{fig:transient dielectric}
\end{figure}

Figure~\ref{fig:transient dielectric} (a and c) shows the calculated $\Delta \Re \epsilon_{xy}$ (i.e., its difference plots) at $t=35$ fs, with respect to the equilibrium state (at $t<0$~fs). To understand these time-dependent change, we further deconvolute the M$_3$ and M$_2$ peaks by projecting them into $3p^{3/2}$ and $3p^{1/2}$ manifolds, respectively. In Fig.~\ref{fig:transient dielectric} (b and d), we observe that in the energy regime below the overlapping region, contribution from the $M_2$ edge is minimal. Furthermore, the M-decomposed $\Delta  \Re \epsilon_{xy}$ is dominantly negative below the M-peak itself, such that overall $\Delta  \Re \epsilon_{xy}$ (in Fig.~\ref{fig:transient dielectric} (a and b)) is dominantly negative at energies below the M$_3$ peak. This suggests that these changes arise from spin flips and the depopulation of the occupied states. Most notably, the overall $\Delta \epsilon_{xy}$ is markedly pronounced in the overlap regime (Fig.~\ref{fig:transient dielectric} (a and c)), exhibiting pronounced oscillatory behaviors, with the M-decomposed $\Delta \epsilon_{xy}$ showing sign changes as well. 

\begin{figure}
    \centering
    \includegraphics[width=\linewidth]{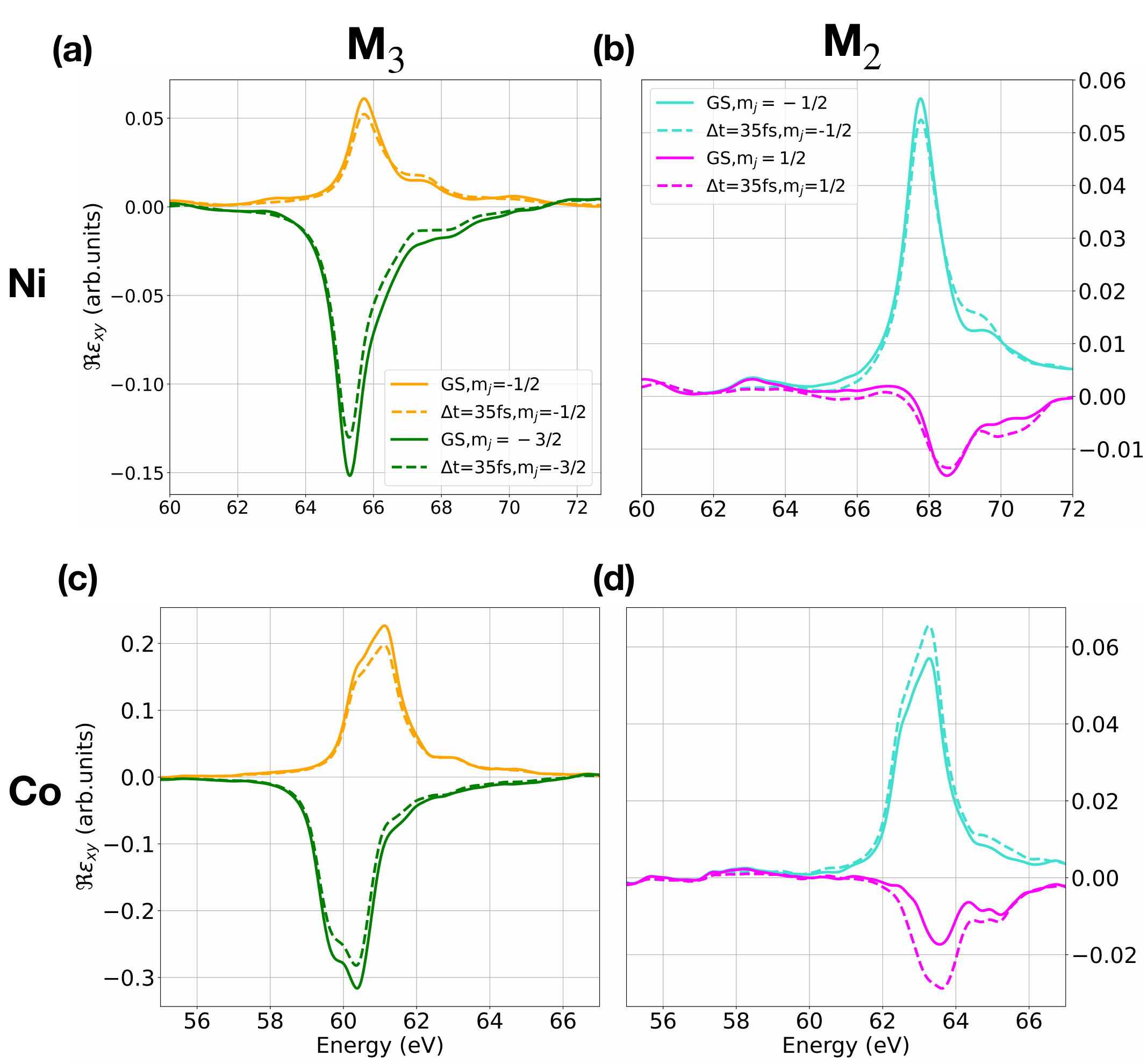}
    \caption{Separation of contributions to the dielectric tensor at the M$_3$ and M$_2$ edges in the ground state (solid lines) and at 35 fs (dashed lines). Panel (a) displays the $m_j=\frac{3}{2}$ and $m_j=-\frac{1}{2}$ states in the $3p^{3/2}$  manifold, and panel (b) shows the $m_j=\frac{1}{2}$ and $m_j=-\frac{1}{2}$ states in the $3p^{1/2}$  manifold for Ni in the ground state and at 35 fs. Panels (c) and (d) show the corresponding states for Co.  }
    \label{fig:ter12stateresolved}
\end{figure}

To elucidate the variations of the dielectric tensor as a function of probe energy, we isolate the constituent $m_j=-\frac{3}{2},-\frac{1}{2}$ contributions at the M$_3$ edge (Fig.~\ref{fig:ter12stateresolved} (a and c)) and the constituent $m_j=-\frac{1}{2},\frac{1}{2}$ contributions at the M$_2$ edge (Fig.~\ref{fig:ter12stateresolved} (b and d)). At the M$_3$ edge of Ni and Co, we see that $35$~fs after the onset of excitation, the magnitudes of the contributions from both $m_j=-\frac{3}{2}$ and $m_j=-\frac{1}{2}$ states have decreased (Fig.~\ref{fig:ter12stateresolved} (a and c)). However, the sign of the $m_j=-\frac{1}{2}$ state has remained positive, while the sign of the $m_j=-\frac{3}{2}$ state has remained negative. The superposition of the individual $m_j$-contributions leads to a partial cancellation of their effects in the overall change of the off-diagonal dielectric tensor component, $\Delta \epsilon_{xy}$ (see Fig.~\ref{fig:transient dielectric} (a and c)). Similarly, at the M$_2$ edge, the superposition of the $m_j = \frac{1}{2}$ and $m_j = -\frac{1}{2}$ states leads to a reduction in the magnitude of the overall peak compared to the individual $m_j$ contributions (see Fig.~\ref{fig:ter12stateresolved}(b) and (d)). Moreover, the high-energy tail of the M$_3$ peak extends above the M$_2$ peak (see Fig.~\ref{fig:ter12stateresolved}(a) and (c)), further complicating the interpretation of the TR-TMOKE signal measured at the M$_2$ edge.

Finally, the response is material specific. Despite both Ni and Co experiencing overall decrease in magnetization after 35~fs, the response of the constituent $m_j$ contributions are qualitatively different  especially at the M$_2$ peak. For Ni, the magnitudes of both the $m_j = \frac{1}{2}, -\frac{1}{2}$ contributions have decreased (Fig.~\ref{fig:ter12stateresolved}b), whereas those of Co has increased (Fig.~\ref{fig:ter12stateresolved}d). In addition to this, due to their partial cancellation, $\Delta \Re \epsilon_{xy}$ at the M$_2$ peak is completely different for both Ni and Co (Fig.~\ref{fig:transient dielectric}(a and c)). The change in Co is also much larger in magnitude than for Ni (Fig. \ref{fig:ter12stateresolved} (b and d)). This is likely to be due to the availability of more empty conduction states near the Fermi level in Co compared to Ni (see Fig.~\ref{fig:schematic}a), since the core states have nearly identical spin-up and spin-down characteristics (see Fig. \textcolor{blue}{S1} in the SM).

A key unresolved question in the field of ultrafast magnetization dynamics is whether the asymmetry observed in TR-TMOKE experiments directly correlates with changes in magnetization. Our findings underscore the central message of this work: to accurately determine whether a material has undergone magnetization or demagnetization using the TR-TMOKE technique, it is insufficient to merely observe that the TMOKE signal has   increased or decreased at an arbitrary, unspecified probe energy. This is because the TMOKE response is a complex function of the probe energy and does not exhibit a straightforward relationship across different energies. 

Quantum mechanics selection rules, modulated by Clebsch-Gordan coefficients, imprint distinct signatures in the dielectric tensor. The SOC in valence and conduction electrons enables spin flips, while in core states it mixes spin components and splits the $3p^{3/2}$ and $3p^{1/2}$ states by 1–3 eV.  The overlap between  the M$_3$ and M$_2$  edges further complicates the interpretation, as dynamics at one edge inevitably affect the other. To ensure more accurate  interpretations of TR-TMOKE signals, we suggest three key steps. First, by comparing experiment and theory, a comprehensive assessment of the critical overlap between the M$_3$ and M$_2$ edges can be made, and contributions from the overlap to the M$_2$ edge can be disentangled.  Second, theoretical analysis should carefully resolve the contributions from each core state, with particular attention to the characteristics of their spin projections. Third, TR-TMOKE experiments should be conducted at multiple probing energies \cite{ryan2023optically}, especially in regions without spectral overlap. 
This is especially important for magnetic alloys, where optical intersite spin transfer is anticipated~\cite{dewhurst2018laser}. For Ni and Co, based on our analysis, this corresponds to the range of energies of  the M$_3$ edge.

\section{Acknowledgements}
O.G. acknowledges the Swedish Research Council (VR) grant 2019-03901 and European Research Council, Synergy Grant 854843-FASTCORR for funding. 
O.E. acknowledges support by the Swedish Research Council (VR), the European Research Council (854843-FASTCORR), eSSENCE and STandUP. O.E. acknowledge support from the Wallenberg Initiative Materials Science for Sustainability (WISE) funded by the Knut and Alice Wallenberg Foundation (KAW). 
The computations were enabled by resources provided by the National Academic Infrastructure for Supercomputing in Sweden (NAISS) and the Swedish National Infrastructure for Computing (SNIC) at NSC and PDC and Uppmax partially funded by the Swedish Research Council through grant agreements No. 2018-05973. and No. 2022-06725 and No. 2018-05973.


%

%

\clearpage
\onecolumngrid
\appendix
\section*{Supplementary Materials}
\setcounter{section}{0}
\setcounter{figure}{0}
\setcounter{table}{0}
\renewcommand{\thefigure}{S\arabic{figure}}
\renewcommand{\thetable}{S\arabic{table}}
\renewcommand{\thesection}{S\arabic{section}}

%
%
%


\title{ Supplementary Materials for: How quantum selection rules influence the magneto-optical effects of driven, ultrafast magnetization dynamics}


\author{Mohamed F. Elhanoty}
\email{mohamed.elhanoty@physics.uu.se}
 \affiliation{Division of Materials Theory, Department of Physics and Astronomy, Uppsala University, Box-516, SE 75120, Sweden}


\author{Olle Eriksson}
\affiliation{Division of Materials Theory, Department of Physics and Astronomy, Uppsala University, Box-516, SE 75120, Sweden}
\affiliation{Wallenberg Initiative Materials Science for Sustainability (WISE), Uppsala University, 75121 Uppsala, Sweden }

\author{Chin Shen Ong}
 \affiliation{Division of Materials Theory, Department of Physics and Astronomy, Uppsala University, Box-516, SE 75120, Sweden}

\author{Oscar Gr\aa n\"as}
\email{Corresponding Author: oscar.granas@physics.uu.se}
\affiliation{Division of Materials Theory, Department of Physics and Astronomy, Uppsala University, Box-516, SE 75120, Sweden}

\date{\today}


\maketitle

\section{Methods}
  This supplementary document provides detailed insights into the methodologies and results supporting the main findings. The reader will find an overview of the theoretical and computational framework used in this study, including Time-Dependent Density Functional Theory (TDDFT) and linear response calculations. Additionally, we discuss the specific numerical details and simulation parameters relevant to our approach.

\subsection{TDDFT}
\renewcommand{\thefigure}{S\arabic{figure}}
\setcounter{figure}{0}  
Through the one to one mapping between the time dependent density $\mathrm{n(\mathbf{r},t)}$ and the external potential $\mathrm{V_{ext}(\mathbf{r},t)}$, Runge and Gross extended the \textit{ab initio} ground state Kohn Sham density functional theory to the time domain~\cite{runge1984density}. This mapping allows for a fully interacting system to be represented by  an equivalent non-interacting one with a time dependent Kohn-Sham (KS) effective potential $v_s$($\mathbf{r}$,t), that produces the same density as the fully interacting system.
 The  KS effective potential, $v_\mathrm{s}(\mathbf{r},t)$,  is a sum of three terms $v_s(\mathbf{r},t)=v_{\mathrm{ext}}(\mathbf{r},t)+v_{\mathrm{H}}(\mathbf{r},t)+v_{\mathrm{xc}}(\mathbf{r},t)$, where $v_{\mathrm{ext}}(\mathbf{r},t)$ is the external potential, $v_{\mathrm{H}}(\mathbf{r},t)$ is the Hartree potential, and  $v_{\mathrm{xc}}(\mathbf{r},t)$ is the exchange-correlation (XC) potential.  The TDKS  can be written as:
\begin{equation}
\begin{split}
    \bigg[\frac{1}{2}{\big (} -i\nabla + \frac{1}{c} \mathbf{A}_{\mathrm{ext}}(t){\big )}^2+v_s(\mathbf{r},t)
    +\frac{1}{2c}\sigma \cdot \mathbf{B}_s(\mathbf{r},t)+ 
    \frac{1}{4c^2} \sigma \cdot {\big (}\nabla v_s(\mathbf{r},t)\times -i\nabla{\big )}
    \bigg] \psi_i(\mathbf{r},t)=\frac{\partial \psi_i(\mathbf{r},t)}{\partial t},
\end{split}
  \label{eq:TDSERG}
\end{equation}
 where c is the speed of light, $\sigma$ is the Pauli matrix, and $\mathbf{B}_s(\mathbf{r},t)$ is the effective KS magnetic field ($\mathbf{B}_s(\mathbf{r},t)=\mathbf{B}_{\mathrm{ext}}(t)+\mathbf{B}_{\mathrm{xc}}(\mathbf{r},t)$, where $\mathbf{B}_{\mathrm{ext}}(t)$ is the magnetic field of the external laser pulse and $\mathbf{B}_\mathrm{xc}(\mathbf{r},t)$ is the xc induced exchange splitting. The last term in Eq. \ref{eq:TDSERG} is the SOC term
 and $\psi_i(\mathbf{r},t)$ is two component Pauli spinor. If the material is driven out of equilibrium by an  external laser, the dipole approximation for the vector potential $\mathrm{\mathbf{A}_{ext}(t)}$ of the pump laser is usually used for wavelengths much larger than the lattice constant.  The atomic units (with $\hbar=e=$m=1) are adopted in Eq. \ref{eq:TDSERG}.

 \subsection{Linear response}
  The  mixed scheme between the time evolution of Eq. \ref{eq:TDSERG} and the linear response introduced by \textit{Dewhurst et al}~\cite{smdewhurst2020element}  is employed but with some modifications presented in Ref.~\cite{smryan2023optically,smlojewski2023interplay} to calculate the response function at a 35 fs time.  In our approach, full transient quantities from the time evolution (TE) of Eq. \ref{eq:TDSERG} used to compute the KS response function in Eq. \ref{eq:kh_chi}, incorporating the effects of the pump laser on both excitation energies and KS states. This contrasts with the approach in Ref. \cite{smdewhurst2020element}, where the excitation energies and KS states are taken from the ground state. We show this contrast in Fig. \ref{fig:compare_mo_sang}.

  The non interacting (KS) response function can be written as \cite{petersilka1996excitation}:

\begin{equation}
    \chi^0(\mathbf{r}, \mathbf{r'}, \omega, t') = \lim_{\eta \rightarrow 0} \sum_{i=1}^\infty \sum_{j=1}^\infty
    (n_i(t') - n_j(t')) 
    \times \frac{\psi_i^{\ast}(\mathbf{r}, t') \psi_j(\mathbf{r}, t')
    \psi_i(\mathbf{r'}, t') \psi_j^{\ast}(\mathbf{r'}, t')}
    {\omega - \epsilon_j(t') + \epsilon_i(t') + i\eta},
    \label{eq:kh_chi}
\end{equation}
where \( \psi_i(\mathbf{r}, t') \) is the KS orbital at the given time step \( t' \), \( \epsilon_i(t') \) is the transient KS eigenvalue, and \( n_i(t') \) is the occupation number at the given time step \( t' \). 

The interacting response function \( \chi \) is related to the non-interacting one \( \chi^0 \) through the following Dyson equation:

\begin{equation}
    \chi(\mathbf{r}, \mathbf{r'}, \omega, t') = \chi^0(\mathbf{r}, \mathbf{r'}, \omega, t') 
    + \chi^0(\mathbf{r}, \mathbf{r'}, \omega, t') \, (\nu + f_{xc}(\omega, t')) \, \chi(\mathbf{r}, \mathbf{r'}, \omega, t')
\end{equation}

where \( \nu \) is the bare Coulomb interaction and \( f_{xc}(\omega, t') \) is the time-dependent exchange-correlation kernel.

The dielectric tensor \( \epsilon(\omega, t') \) is calculated using the response function according to the following equation:
\begin{equation}
    \epsilon_{ij}^{-1}(\omega, t') = \delta_{ij} + \nu \chi_{ij}(\omega, t') \label{eq:epsinv}.
\end{equation}
This formulation accounts for the time dependence of the occupation numbers, KS eigenvalues, orbitals, and the exchange-correlation kernel, making the response function applicable at specific instantaneous time steps after the evolution of the KS equations.

\subsection{Numerical Details}
For each system, we applied a sine pulse with a carrier energy of 1.55~eV, modulated by a gaussian envelope that has a full width at half maximum (FWHM) duration of 35~fs. This corresponds to a carrier wavelength of 800~nm and a fluence of 12~mJ/cm$^2$. The pulse is turned on at $t=0$~fs where $t$ is defined as the time since the onset of the time-dependent external optical field.  These are parameters typically used in experimental studies of ultrafast magnetization dynamics.  The transient linear response calculations were performed as detailed in Ref.~\cite{smryan2023optically,smlojewski2023interplay}. The time evolution of Eq. \ref{eq:TDSERG} and the linear response calculation are performed using the full potential basis ELK code~\cite{smelk}.


\subsection{Transverse MOKE}
 
Magneto-optical Kerr effect (MOKE) refers to the phenomenon where the polarization state of light is altered upon reflection from a magnetized material. When linearly polarized light reflects off a magnetized surface, the interaction between the light's electric field and the material's magnetization ($M$) causes a rotation in the plane of polarization. Transverse MOKE (TMOKE) refers to a specific configuration of MOKE where $M$ lies in the plane of the surface and is perpendicular to the plane of incidence.

In magneto-optical materials, $\epsilon$ is  anisotropic and has off-diagonal components ($\epsilon_{xy}$) due to the magnetic order (i.e., $M$) and SOC. The real part of $\epsilon_{xy}$  (i.e., $\Re \epsilon_{xy}$) is related to the refractive index, $n$ and determines the phase velocity of light. The sign of the peaks of $\Re \epsilon_{xy}$ corresponds to the the direction of the rotation of the polarization plane that is directly related to the sign of $M$.

The Kerr effect is intrinsically related to the dielectric tensor, $\epsilon$, of the material. Here, $\epsilon$ represents the linear response that characterizes the relationship between the electric displacement field, $\mathbf{D}$, and the applied electric field, $\mathbf{E}$. In magneto-optical materials, $\epsilon$ becomes anisotropic and includes off-diagonal components ($\epsilon_{xy}$) due to magnetic order (i.e., $M$) and spin-orbit coupling. The real part of $\epsilon_{xy}$ (i.e., $\Re \epsilon_{xy}$) is linked to the refractive index, $n$, and governs the phase velocity of light. For more details, see, for example, Ref.~\cite{smoppeneer2001magneto}.

The experimental observable  in the TMOKE experiment is the magnetic asymmetry that is defined as the difference between the reflectances for the two
 magnetization directions $I_{\pm}$
  normalized to their sum:
 \begin{equation}
   A=\frac{I_{+}-I_{-}}{I_{+}+I_{-}}
 \end{equation}
 where
 \begin{equation}
     I_{\pm}=|I_{0}|^2+I_m\epsilon_{xy}\pm \Re\{2I_0 I_m \epsilon_{xy}\}
 \end{equation}
 where $I_0=\frac{n \cos{\theta_i}-\cos{\theta_t}}{n\cos \theta_i + \cos \theta_t}$ and $I_m=\frac{\sin 2\theta_i}{n^2(n\cos\theta_i+\cos \theta_t)}$. The $\theta_i$ and $\theta_t$ are the incident and the transmission angles respectively and  n is the refractive index.

\subsection{Spin Projections of 3p Core States }
\begin{figure}
    \centering
    \includegraphics[width=\linewidth]{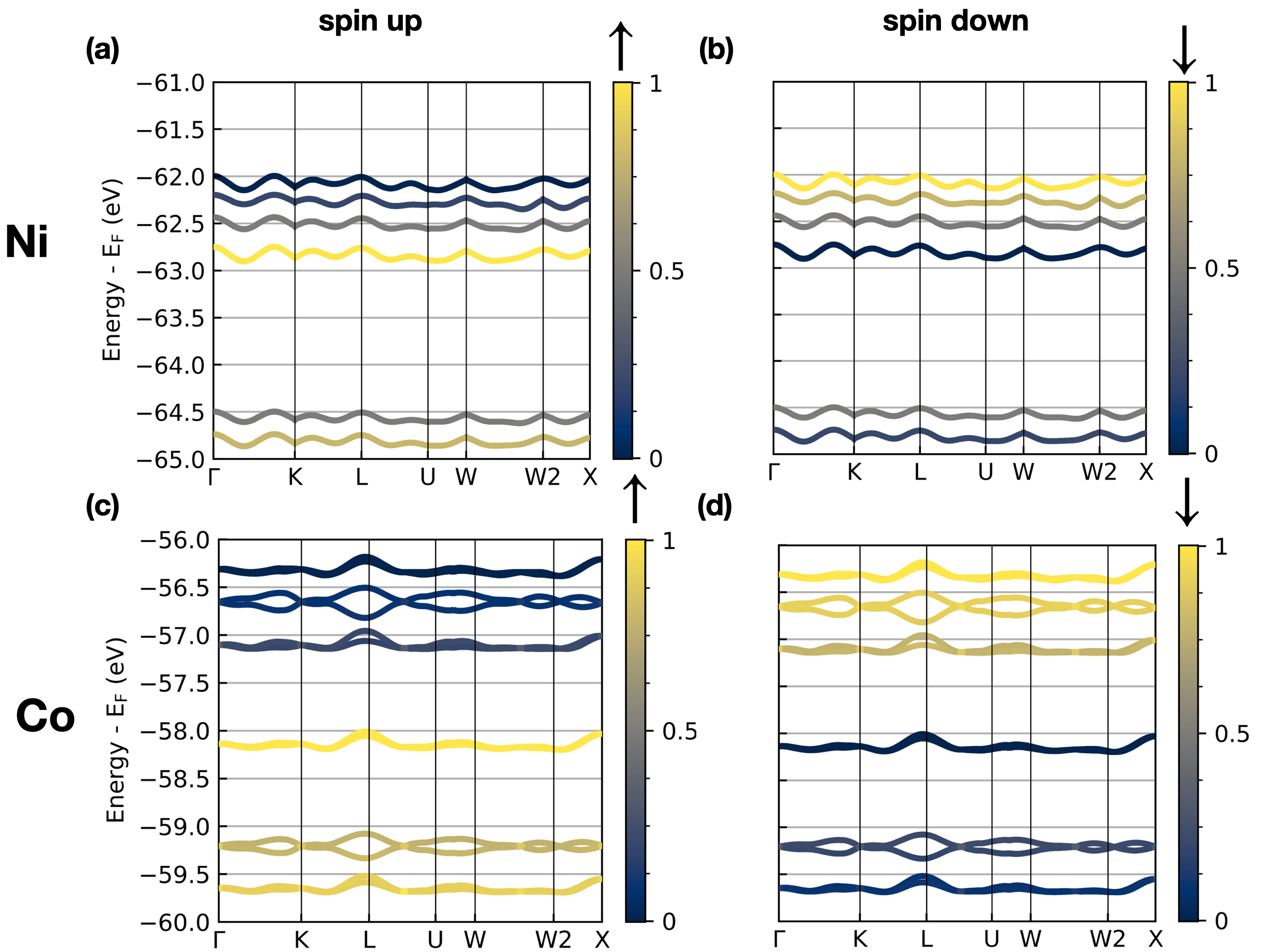}
    \caption{Spin-up and spin-down projections of the ground state band structure for the 3$p$ states. Panels (a) and (b) depict the spin-up and spin-down projections for Ni, respectively, and panels (c) and (d) show the corresponding projections for Co.}
    \label{fig:GS_Band_structure}
\end{figure}

In atomic systems, the choice between Russell-Saunders  coupling (LS) and \(jj\) coupling depends on the competition between the SOC and electrostatic interactions~\cite{condon1935theory, bethe2013quantum}. For valence electrons in lighter atoms,  the LS coupling is more suitable due to weaker SOC~\cite{bransden2003physics}. However, for core electrons, where SOC is stronger, the \(jj\) coupling scheme becomes more appropriate and provides a more diagonal representation~\cite{bransden2003physics}. The presence of relatively stronger SOC for the core states leads to mixing the spin up and spin down components. In addition, the spin angular momentum is no longer a good quantum number due to SOC. Fig.~\ref{fig:GS_Band_structure} depicts the relative weights of their bands in the spin-up and spin-down projections for Co and Ni. The mixing of spin components in the core states plays a key role in explaining the energy dependence of both static and transient signals as discussed in detail below.
To elucidate the impact of the spin mixing in the core states on both static and transient signals, Fig.~\ref{fig:GS_Band_structure} illustrates the relative weights of their bands in the spin-up and spin-down projections.
The $3p^{1/2}$  and $3p^{3/2}$  states consist of two and four states, respectively, each exhibiting varying contributions of spin-up and spin-down projections, as shown in Fig.~\ref{fig:GS_Band_structure}. Note that since Co in the hcp phase has two atoms per cell, Fig.~\ref{fig:GS_Band_structure} shows twice as many states compared to Ni, with only one atom per unit cell.
In the $3p^{1/2}$  manifold, the state with $m_j=\frac{1}{2}$ for both elements has its primary contribution from spin-up, while the state with $m_j=-\frac{1}{2}$ has its main contribution from spin-down. Given that most of the empty states above the Fermi level, available for optical excitation, exhibit spin-down character (see Fig. \textcolor{blue}{1}a in the main paper), the optical transitions with spin-down character contribute significantly to the final signal, resulting in a more intense contribution from the $m_j=-\frac{1}{2}$ state than the $m_j=\frac{1}{2}$ state, see Fig.2b and d in the main paper. 

Similarly, the $m_j=-\frac{3}{2}$ state is primarily composed of spin-down contributions, while the $m_j=-\frac{1}{2}$ and $m_j=\frac{1}{2}$ states are mixtures of both spin-up and spin-down components. Therefore, the $m_j=-\frac{3}{2}$ state manifests a strong negative contribution, while the $m_j=-\frac{1}{2}$ and $m_j=\frac{1}{2}$ states produce smaller positive contributions, leading to a net overall negative contribution at the M$_3$ edge, see Fig. \textcolor{blue}{2}  (a and c) in the main paper.
\section{Comparison between Real-Time Approach and Rigid Band Structure}
In this work, we employ a mixed scheme incorporating the time evolution (TE) of Eq.~\ref{eq:TDSERG} with linear response to calculate the dielectric tensor response at 35 fs. Unlike previous methods, which rely on a rigid band structure with ground-state (GS) excitation energies and Kohn-Sham (KS) states, our approach utilizes full transient quantities that dynamically adjust under the influence of the pump laser. This distinction enables our method to capture dynamical changes in the excitation energies and KS states, accommodating transient changes that would otherwise be lost if KS states were fixed to their GS configurations.

\begin{figure}[H]
    \centering
    \includegraphics[width=\linewidth]{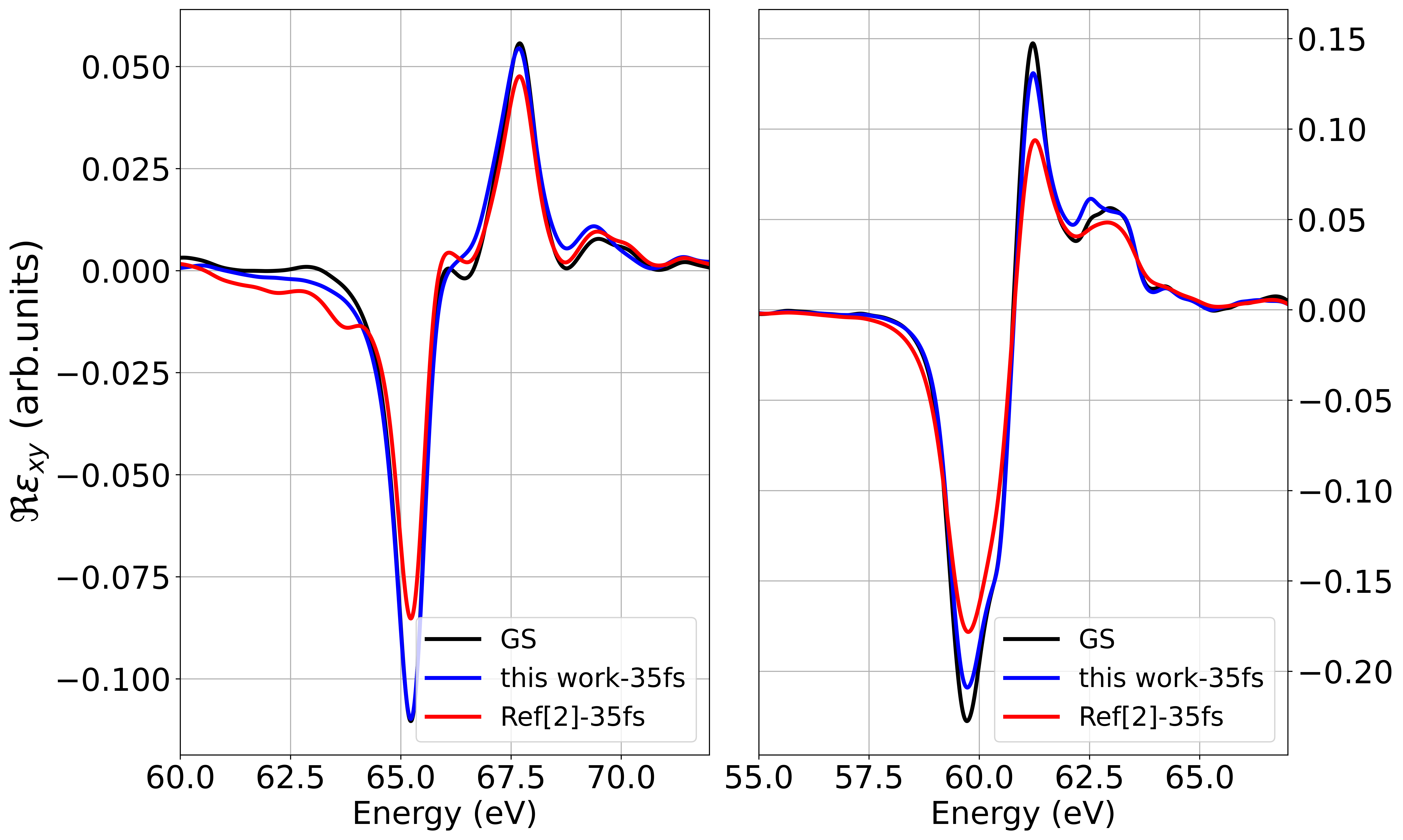}
    \caption{Comparison of the calculated dielectric tensor response at 35 fs using the full transient quantities from the time evolution of Eq.~\ref{eq:TDSERG} (this work) and the rigid band structure method used in Ref.~\cite{smdewhurst2020element}.}
    \label{fig:compare_mo_sang}
\end{figure}

Figure~\ref{fig:compare_mo_sang} illustrates the contrast between our approach and the rigid band structure method from Ref.~\cite{smdewhurst2020element}. By maintaining GS excitation energies and KS states, the rigid band structure method limits the exploration of selection rule effects on ultrafast dynamics. In particular, this restriction prevents any real-time adjustments in the dipole matrix elements, which govern optical transitions from core to valence states. Our approach, on the other hand,  incorporates these transient variations, allowing the dipole matrix elements to reflect the symmetry changes induced by the pump laser. This flexibility results in a dielectric response that more closely aligns with experimental observables, including high-resolution time-resolved XAS and TMOKE measurements \cite{smlojewski2023interplay,smryan2023optically}.

\begin{thebibliography}{24}%
\makeatletter
\providecommand \@ifxundefined [1]{%
 \@ifx{#1\undefined}
}%
\providecommand \@ifnum [1]{%
 \ifnum #1\expandafter \@firstoftwo
 \else \expandafter \@secondoftwo
 \fi
}%
\providecommand \@ifx [1]{%
 \ifx #1\expandafter \@firstoftwo
 \else \expandafter \@secondoftwo
 \fi
}%
\providecommand \natexlab [1]{#1}%
\providecommand \enquote  [1]{``#1''}%
\providecommand \bibnamefont  [1]{#1}%
\providecommand \bibfnamefont [1]{#1}%
\providecommand \citenamefont [1]{#1}%
\providecommand \href@noop [0]{\@secondoftwo}%
\providecommand \href [0]{\begingroup \@sanitize@url \@href}%
\providecommand \@href[1]{\@@startlink{#1}\@@href}%
\providecommand \@@href[1]{\endgroup#1\@@endlink}%
\providecommand \@sanitize@url [0]{\catcode `\\12\catcode `\$12\catcode
  `\&12\catcode `\#12\catcode `\^12\catcode `\_12\catcode `\%12\relax}%
\providecommand \@@startlink[1]{}%
\providecommand \@@endlink[0]{}%
\providecommand \url  [0]{\begingroup\@sanitize@url \@url }%
\providecommand \@url [1]{\endgroup\@href {#1}{\urlprefix }}%
\providecommand \urlprefix  [0]{URL }%
\providecommand \Eprint [0]{\href }%
\providecommand \doibase [0]{http://dx.doi.org/}%
\providecommand \selectlanguage [0]{\@gobble}%
\providecommand \bibinfo  [0]{\@secondoftwo}%
\providecommand \bibfield  [0]{\@secondoftwo}%
\providecommand \translation [1]{[#1]}%
\providecommand \BibitemOpen [0]{}%
\providecommand \bibitemStop [0]{}%
\providecommand \bibitemNoStop [0]{.\EOS\space}%
\providecommand \EOS [0]{\spacefactor3000\relax}%
\providecommand \BibitemShut  [1]{\csname bibitem#1\endcsname}%
\let\auto@bib@innerbib\@empty
\bibitem [{\citenamefont {Bhushan}(2023)}]{bhushan2023current}%
  \BibitemOpen
  \bibfield  {author} {\bibinfo {author} {\bibfnamefont {B.}~\bibnamefont
  {Bhushan}},\ }\bibfield  {title} {\enquote {\bibinfo {title} {Current status
  and outlook of magnetic data storage devices},}\ }\href@noop {} {\bibfield
  {journal} {\bibinfo  {journal} {Microsystem Technologies}\ }\textbf {\bibinfo
  {volume} {29}},\ \bibinfo {pages} {1529--1546} (\bibinfo {year}
  {2023})}\BibitemShut {NoStop}%
\bibitem [{\citenamefont {Dieny}\ \emph {et~al.}(2020)\citenamefont {Dieny},
  \citenamefont {Prejbeanu}, \citenamefont {Garello}, \citenamefont
  {Gambardella}, \citenamefont {Freitas}, \citenamefont {Lehndorff},
  \citenamefont {Raberg}, \citenamefont {Ebels}, \citenamefont {Demokritov},
  \citenamefont {Akerman} \emph {et~al.}}]{dieny2020opportunities}%
  \BibitemOpen
  \bibfield  {author} {\bibinfo {author} {\bibfnamefont {B.}~\bibnamefont
  {Dieny}}, \bibinfo {author} {\bibfnamefont {I.~L.}\ \bibnamefont
  {Prejbeanu}}, \bibinfo {author} {\bibfnamefont {K.}~\bibnamefont {Garello}},
  \bibinfo {author} {\bibfnamefont {P.}~\bibnamefont {Gambardella}}, \bibinfo
  {author} {\bibfnamefont {P.}~\bibnamefont {Freitas}}, \bibinfo {author}
  {\bibfnamefont {R.}~\bibnamefont {Lehndorff}}, \bibinfo {author}
  {\bibfnamefont {W.}~\bibnamefont {Raberg}}, \bibinfo {author} {\bibfnamefont
  {U.}~\bibnamefont {Ebels}}, \bibinfo {author} {\bibfnamefont {S.~O.}\
  \bibnamefont {Demokritov}}, \bibinfo {author} {\bibfnamefont
  {J.}~\bibnamefont {Akerman}},  \emph {et~al.},\ }\bibfield  {title} {\enquote
  {\bibinfo {title} {Opportunities and challenges for spintronics in the
  microelectronics industry},}\ }\href@noop {} {\bibfield  {journal} {\bibinfo
  {journal} {Nature Electronics}\ }\textbf {\bibinfo {volume} {3}},\ \bibinfo
  {pages} {446--459} (\bibinfo {year} {2020})}\BibitemShut {NoStop}%
\bibitem [{\citenamefont {Hirohata}\ \emph {et~al.}(2020)\citenamefont
  {Hirohata}, \citenamefont {Yamada}, \citenamefont {Nakatani}, \citenamefont
  {Prejbeanu}, \citenamefont {Diény}, \citenamefont {Pirro},\ and\
  \citenamefont {Hillebrands}}]{hirohata2020review}%
  \BibitemOpen
  \bibfield  {author} {\bibinfo {author} {\bibfnamefont {A.}~\bibnamefont
  {Hirohata}}, \bibinfo {author} {\bibfnamefont {K.}~\bibnamefont {Yamada}},
  \bibinfo {author} {\bibfnamefont {Y.}~\bibnamefont {Nakatani}}, \bibinfo
  {author} {\bibfnamefont {I.-L.}\ \bibnamefont {Prejbeanu}}, \bibinfo {author}
  {\bibfnamefont {B.}~\bibnamefont {Diény}}, \bibinfo {author} {\bibfnamefont
  {P.}~\bibnamefont {Pirro}}, \ and\ \bibinfo {author} {\bibfnamefont
  {B.}~\bibnamefont {Hillebrands}},\ }\bibfield  {title} {\enquote {\bibinfo
  {title} {Review on spintronics: Principles and device applications},}\ }\href
  {\doibase https://doi.org/10.1016/j.jmmm.2020.166711} {\bibfield  {journal}
  {\bibinfo  {journal} {Journal of Magnetism and Magnetic Materials}\ }\textbf
  {\bibinfo {volume} {509}},\ \bibinfo {pages} {166711} (\bibinfo {year}
  {2020})}\BibitemShut {NoStop}%
\bibitem [{\citenamefont {Shirozhan}\ \emph {et~al.}(2024)\citenamefont
  {Shirozhan}, \citenamefont {Mondal}, \citenamefont {Grósz}, \citenamefont
  {Nagyillés}, \citenamefont {Farkas}, \citenamefont {Nayak}, \citenamefont
  {Ahmed}, \citenamefont {Dey}, \citenamefont {Marco}, \citenamefont
  {Nelissen}, \citenamefont {Kiss}, \citenamefont {Oldal}, \citenamefont
  {Csizmadia}, \citenamefont {Filus}, \citenamefont {Marco}, \citenamefont
  {Madas}, \citenamefont {Kahaly}, \citenamefont {Charalambidis}, \citenamefont
  {Tzallas}, \citenamefont {Appi}, \citenamefont {Weissenbilder}, \citenamefont
  {Eng-Johnsson}, \citenamefont {L’Huillier}, \citenamefont {Diveki},
  \citenamefont {Major}, \citenamefont {Varjú},\ and\ \citenamefont
  {Kahaly}}]{shirozhan2024high}%
  \BibitemOpen
  \bibfield  {author} {\bibinfo {author} {\bibfnamefont {M.}~\bibnamefont
  {Shirozhan}}, \bibinfo {author} {\bibfnamefont {S.}~\bibnamefont {Mondal}},
  \bibinfo {author} {\bibfnamefont {T.}~\bibnamefont {Grósz}}, \bibinfo
  {author} {\bibfnamefont {B.}~\bibnamefont {Nagyillés}}, \bibinfo {author}
  {\bibfnamefont {B.}~\bibnamefont {Farkas}}, \bibinfo {author} {\bibfnamefont
  {A.}~\bibnamefont {Nayak}}, \bibinfo {author} {\bibfnamefont
  {N.}~\bibnamefont {Ahmed}}, \bibinfo {author} {\bibfnamefont
  {I.}~\bibnamefont {Dey}}, \bibinfo {author} {\bibfnamefont {S.~C.~D.}\
  \bibnamefont {Marco}}, \bibinfo {author} {\bibfnamefont {K.}~\bibnamefont
  {Nelissen}}, \bibinfo {author} {\bibfnamefont {M.}~\bibnamefont {Kiss}},
  \bibinfo {author} {\bibfnamefont {L.~G.}\ \bibnamefont {Oldal}}, \bibinfo
  {author} {\bibfnamefont {T.}~\bibnamefont {Csizmadia}}, \bibinfo {author}
  {\bibfnamefont {Z.}~\bibnamefont {Filus}}, \bibinfo {author} {\bibfnamefont
  {M.~D.}\ \bibnamefont {Marco}}, \bibinfo {author} {\bibfnamefont
  {S.}~\bibnamefont {Madas}}, \bibinfo {author} {\bibfnamefont {M.~U.}\
  \bibnamefont {Kahaly}}, \bibinfo {author} {\bibfnamefont {D.}~\bibnamefont
  {Charalambidis}}, \bibinfo {author} {\bibfnamefont {P.}~\bibnamefont
  {Tzallas}}, \bibinfo {author} {\bibfnamefont {E.}~\bibnamefont {Appi}},
  \bibinfo {author} {\bibfnamefont {R.}~\bibnamefont {Weissenbilder}}, \bibinfo
  {author} {\bibfnamefont {P.}~\bibnamefont {Eng-Johnsson}}, \bibinfo {author}
  {\bibfnamefont {A.}~\bibnamefont {L’Huillier}}, \bibinfo {author}
  {\bibfnamefont {Z.}~\bibnamefont {Diveki}}, \bibinfo {author} {\bibfnamefont
  {B.}~\bibnamefont {Major}}, \bibinfo {author} {\bibfnamefont
  {K.}~\bibnamefont {Varjú}}, \ and\ \bibinfo {author} {\bibfnamefont
  {S.}~\bibnamefont {Kahaly}},\ }\bibfield  {title} {\enquote {\bibinfo {title}
  {High-repetition-rate attosecond extreme ultraviolet beamlines at eli alps
  for studying ultrafast phenomena},}\ }\href {\doibase
  10.34133/ultrafastscience.0067} {\bibfield  {journal} {\bibinfo  {journal}
  {Ultrafast Science}\ }\textbf {\bibinfo {volume} {4}},\ \bibinfo {pages}
  {0067} (\bibinfo {year} {2024})},\ \Eprint
  {http://arxiv.org/abs/https://spj.science.org/doi/pdf/10.34133/ultrafastscience.0067}
  {https://spj.science.org/doi/pdf/10.34133/ultrafastscience.0067} \BibitemShut
  {NoStop}%
\bibitem [{\citenamefont {Kirilyuk}\ \emph {et~al.}(2010)\citenamefont
  {Kirilyuk}, \citenamefont {Kimel},\ and\ \citenamefont
  {Rasing}}]{kirilyuk2010ultrafast}%
  \BibitemOpen
  \bibfield  {author} {\bibinfo {author} {\bibfnamefont {A.}~\bibnamefont
  {Kirilyuk}}, \bibinfo {author} {\bibfnamefont {A.~V.}\ \bibnamefont {Kimel}},
  \ and\ \bibinfo {author} {\bibfnamefont {T.}~\bibnamefont {Rasing}},\
  }\bibfield  {title} {\enquote {\bibinfo {title} {Ultrafast optical
  manipulation of magnetic order},}\ }\href {\doibase
  10.1103/RevModPhys.82.2731} {\bibfield  {journal} {\bibinfo  {journal} {Rev.
  Mod. Phys.}\ }\textbf {\bibinfo {volume} {82}},\ \bibinfo {pages}
  {2731--2784} (\bibinfo {year} {2010})}\BibitemShut {NoStop}%
\bibitem [{\citenamefont {Wang}\ and\ \citenamefont
  {Liu}(2020)}]{wang2020ultrafast}%
  \BibitemOpen
  \bibfield  {author} {\bibinfo {author} {\bibfnamefont {C.}~\bibnamefont
  {Wang}}\ and\ \bibinfo {author} {\bibfnamefont {Y.}~\bibnamefont {Liu}},\
  }\bibfield  {title} {\enquote {\bibinfo {title} {Ultrafast optical
  manipulation of magnetic order in ferromagnetic materials},}\ }\href@noop {}
  {\bibfield  {journal} {\bibinfo  {journal} {Nano Convergence}\ }\textbf
  {\bibinfo {volume} {7}},\ \bibinfo {pages} {1--16} (\bibinfo {year}
  {2020})}\BibitemShut {NoStop}%
\bibitem [{\citenamefont {Kimel}\ \emph {et~al.}(2022)\citenamefont {Kimel},
  \citenamefont {Zvezdin}, \citenamefont {Sharma}, \citenamefont {Shallcross},
  \citenamefont {de~Sousa}, \citenamefont {García-Martín}, \citenamefont
  {Salvan}, \citenamefont {Hamrle}, \citenamefont {Stejskal}, \citenamefont
  {McCord}, \citenamefont {Tacchi}, \citenamefont {Carlotti}, \citenamefont
  {Gambardella}, \citenamefont {Salis}, \citenamefont {Münzenberg},
  \citenamefont {Schultze}, \citenamefont {Temnov}, \citenamefont {Bychkov},
  \citenamefont {Kotov}, \citenamefont {Maccaferri}, \citenamefont {Ignatyeva},
  \citenamefont {Belotelov}, \citenamefont {Donnelly}, \citenamefont
  {Rodriguez}, \citenamefont {Matsuda}, \citenamefont {Ruchon}, \citenamefont
  {Fanciulli}, \citenamefont {Sacchi}, \citenamefont {Du}, \citenamefont
  {Wang}, \citenamefont {Armitage}, \citenamefont {Schubert}, \citenamefont
  {Darakchieva}, \citenamefont {Liu}, \citenamefont {Huang}, \citenamefont
  {Ding}, \citenamefont {Berger},\ and\ \citenamefont
  {Vavassori}}]{kimel20222022}%
  \BibitemOpen
  \bibfield  {author} {\bibinfo {author} {\bibfnamefont {A.}~\bibnamefont
  {Kimel}}, \bibinfo {author} {\bibfnamefont {A.}~\bibnamefont {Zvezdin}},
  \bibinfo {author} {\bibfnamefont {S.}~\bibnamefont {Sharma}}, \bibinfo
  {author} {\bibfnamefont {S.}~\bibnamefont {Shallcross}}, \bibinfo {author}
  {\bibfnamefont {N.}~\bibnamefont {de~Sousa}}, \bibinfo {author}
  {\bibfnamefont {A.}~\bibnamefont {García-Martín}}, \bibinfo {author}
  {\bibfnamefont {G.}~\bibnamefont {Salvan}}, \bibinfo {author} {\bibfnamefont
  {J.}~\bibnamefont {Hamrle}}, \bibinfo {author} {\bibfnamefont
  {O.}~\bibnamefont {Stejskal}}, \bibinfo {author} {\bibfnamefont
  {J.}~\bibnamefont {McCord}}, \bibinfo {author} {\bibfnamefont
  {S.}~\bibnamefont {Tacchi}}, \bibinfo {author} {\bibfnamefont
  {G.}~\bibnamefont {Carlotti}}, \bibinfo {author} {\bibfnamefont
  {P.}~\bibnamefont {Gambardella}}, \bibinfo {author} {\bibfnamefont
  {G.}~\bibnamefont {Salis}}, \bibinfo {author} {\bibfnamefont
  {M.}~\bibnamefont {Münzenberg}}, \bibinfo {author} {\bibfnamefont
  {M.}~\bibnamefont {Schultze}}, \bibinfo {author} {\bibfnamefont
  {V.}~\bibnamefont {Temnov}}, \bibinfo {author} {\bibfnamefont {I.~V.}\
  \bibnamefont {Bychkov}}, \bibinfo {author} {\bibfnamefont {L.~N.}\
  \bibnamefont {Kotov}}, \bibinfo {author} {\bibfnamefont {N.}~\bibnamefont
  {Maccaferri}}, \bibinfo {author} {\bibfnamefont {D.}~\bibnamefont
  {Ignatyeva}}, \bibinfo {author} {\bibfnamefont {V.}~\bibnamefont
  {Belotelov}}, \bibinfo {author} {\bibfnamefont {C.}~\bibnamefont {Donnelly}},
  \bibinfo {author} {\bibfnamefont {A.~H.}\ \bibnamefont {Rodriguez}}, \bibinfo
  {author} {\bibfnamefont {I.}~\bibnamefont {Matsuda}}, \bibinfo {author}
  {\bibfnamefont {T.}~\bibnamefont {Ruchon}}, \bibinfo {author} {\bibfnamefont
  {M.}~\bibnamefont {Fanciulli}}, \bibinfo {author} {\bibfnamefont
  {M.}~\bibnamefont {Sacchi}}, \bibinfo {author} {\bibfnamefont {C.~R.}\
  \bibnamefont {Du}}, \bibinfo {author} {\bibfnamefont {H.}~\bibnamefont
  {Wang}}, \bibinfo {author} {\bibfnamefont {N.~P.}\ \bibnamefont {Armitage}},
  \bibinfo {author} {\bibfnamefont {M.}~\bibnamefont {Schubert}}, \bibinfo
  {author} {\bibfnamefont {V.}~\bibnamefont {Darakchieva}}, \bibinfo {author}
  {\bibfnamefont {B.}~\bibnamefont {Liu}}, \bibinfo {author} {\bibfnamefont
  {Z.}~\bibnamefont {Huang}}, \bibinfo {author} {\bibfnamefont
  {B.}~\bibnamefont {Ding}}, \bibinfo {author} {\bibfnamefont {A.}~\bibnamefont
  {Berger}}, \ and\ \bibinfo {author} {\bibfnamefont {P.}~\bibnamefont
  {Vavassori}},\ }\bibfield  {title} {\enquote {\bibinfo {title} {The 2022
  magneto-optics roadmap},}\ }\href {\doibase 10.1088/1361-6463/ac8da0}
  {\bibfield  {journal} {\bibinfo  {journal} {Journal of Physics D: Applied
  Physics}\ }\textbf {\bibinfo {volume} {55}},\ \bibinfo {pages} {463003}
  (\bibinfo {year} {2022})}\BibitemShut {NoStop}%
\bibitem [{\citenamefont {Ryan}\ \emph {et~al.}(2023)\citenamefont {Ryan},
  \citenamefont {Johnsen}, \citenamefont {Elhanoty}, \citenamefont {Grafov},
  \citenamefont {Li}, \citenamefont {Delin}, \citenamefont {Markou},
  \citenamefont {Lesne}, \citenamefont {Felser}, \citenamefont {Eriksson},
  \citenamefont {Kapteyn}, \citenamefont {Grånäs},\ and\ \citenamefont
  {Murnane}}]{ryan2023optically}%
  \BibitemOpen
  \bibfield  {author} {\bibinfo {author} {\bibfnamefont {S.~A.}\ \bibnamefont
  {Ryan}}, \bibinfo {author} {\bibfnamefont {P.~C.}\ \bibnamefont {Johnsen}},
  \bibinfo {author} {\bibfnamefont {M.~F.}\ \bibnamefont {Elhanoty}}, \bibinfo
  {author} {\bibfnamefont {A.}~\bibnamefont {Grafov}}, \bibinfo {author}
  {\bibfnamefont {N.}~\bibnamefont {Li}}, \bibinfo {author} {\bibfnamefont
  {A.}~\bibnamefont {Delin}}, \bibinfo {author} {\bibfnamefont
  {A.}~\bibnamefont {Markou}}, \bibinfo {author} {\bibfnamefont
  {E.}~\bibnamefont {Lesne}}, \bibinfo {author} {\bibfnamefont
  {C.}~\bibnamefont {Felser}}, \bibinfo {author} {\bibfnamefont
  {O.}~\bibnamefont {Eriksson}}, \bibinfo {author} {\bibfnamefont {H.~C.}\
  \bibnamefont {Kapteyn}}, \bibinfo {author} {\bibfnamefont {O.}~\bibnamefont
  {Grånäs}}, \ and\ \bibinfo {author} {\bibfnamefont {M.~M.}\ \bibnamefont
  {Murnane}},\ }\bibfield  {title} {\enquote {\bibinfo {title} {Optically
  controlling the competition between spin flips and intersite spin transfer in
  a heusler half-metal on sub–100-fs time scales},}\ }\href {\doibase
  10.1126/sciadv.adi1428} {\bibfield  {journal} {\bibinfo  {journal} {Science
  Advances}\ }\textbf {\bibinfo {volume} {9}},\ \bibinfo {pages} {eadi1428}
  (\bibinfo {year} {2023})},\ \Eprint
  {http://arxiv.org/abs/https://www.science.org/doi/pdf/10.1126/sciadv.adi1428}
  {https://www.science.org/doi/pdf/10.1126/sciadv.adi1428} \BibitemShut
  {NoStop}%
\bibitem [{\citenamefont {Buades}\ \emph {et~al.}(2021)\citenamefont {Buades},
  \citenamefont {Picón}, \citenamefont {Berger}, \citenamefont {León},
  \citenamefont {Di~Palo}, \citenamefont {Cousin}, \citenamefont {Cocchi},
  \citenamefont {Pellegrin}, \citenamefont {Martin}, \citenamefont
  {Mañas-Valero}, \citenamefont {Coronado}, \citenamefont {Danz},
  \citenamefont {Draxl}, \citenamefont {Uemoto}, \citenamefont {Yabana},
  \citenamefont {Schultze}, \citenamefont {Wall}, \citenamefont {Zürch},\ and\
  \citenamefont {Biegert}}]{buades2021attosecond}%
  \BibitemOpen
  \bibfield  {author} {\bibinfo {author} {\bibfnamefont {B.}~\bibnamefont
  {Buades}}, \bibinfo {author} {\bibfnamefont {A.}~\bibnamefont {Picón}},
  \bibinfo {author} {\bibfnamefont {E.}~\bibnamefont {Berger}}, \bibinfo
  {author} {\bibfnamefont {I.}~\bibnamefont {León}}, \bibinfo {author}
  {\bibfnamefont {N.}~\bibnamefont {Di~Palo}}, \bibinfo {author} {\bibfnamefont
  {S.~L.}\ \bibnamefont {Cousin}}, \bibinfo {author} {\bibfnamefont
  {C.}~\bibnamefont {Cocchi}}, \bibinfo {author} {\bibfnamefont
  {E.}~\bibnamefont {Pellegrin}}, \bibinfo {author} {\bibfnamefont {J.~H.}\
  \bibnamefont {Martin}}, \bibinfo {author} {\bibfnamefont {S.}~\bibnamefont
  {Mañas-Valero}}, \bibinfo {author} {\bibfnamefont {E.}~\bibnamefont
  {Coronado}}, \bibinfo {author} {\bibfnamefont {T.}~\bibnamefont {Danz}},
  \bibinfo {author} {\bibfnamefont {C.}~\bibnamefont {Draxl}}, \bibinfo
  {author} {\bibfnamefont {M.}~\bibnamefont {Uemoto}}, \bibinfo {author}
  {\bibfnamefont {K.}~\bibnamefont {Yabana}}, \bibinfo {author} {\bibfnamefont
  {M.}~\bibnamefont {Schultze}}, \bibinfo {author} {\bibfnamefont
  {S.}~\bibnamefont {Wall}}, \bibinfo {author} {\bibfnamefont {M.}~\bibnamefont
  {Zürch}}, \ and\ \bibinfo {author} {\bibfnamefont {J.}~\bibnamefont
  {Biegert}},\ }\bibfield  {title} {\enquote {\bibinfo {title} {{Attosecond
  state-resolved carrier motion in quantum materials probed by soft x-ray
  XANES}},}\ }\href {\doibase 10.1063/5.0020649} {\bibfield  {journal}
  {\bibinfo  {journal} {Applied Physics Reviews}\ }\textbf {\bibinfo {volume}
  {8}},\ \bibinfo {pages} {011408} (\bibinfo {year} {2021})},\ \Eprint
  {http://arxiv.org/abs/https://pubs.aip.org/aip/apr/article-pdf/doi/10.1063/5.0020649/14580075/011408\_1\_online.pdf}
  {https://pubs.aip.org/aip/apr/article-pdf/doi/10.1063/5.0020649/14580075/011408\_1\_online.pdf}
  \BibitemShut {NoStop}%
\bibitem [{\citenamefont {La-O-Vorakiat}\ \emph {et~al.}(2009)\citenamefont
  {La-O-Vorakiat}, \citenamefont {Siemens}, \citenamefont {Murnane},
  \citenamefont {Kapteyn}, \citenamefont {Mathias}, \citenamefont
  {Aeschlimann}, \citenamefont {Grychtol}, \citenamefont {Adam}, \citenamefont
  {Schneider}, \citenamefont {Shaw}, \citenamefont {Nembach},\ and\
  \citenamefont {Silva}}]{la2009ultrafast}%
  \BibitemOpen
  \bibfield  {author} {\bibinfo {author} {\bibfnamefont {C.}~\bibnamefont
  {La-O-Vorakiat}}, \bibinfo {author} {\bibfnamefont {M.}~\bibnamefont
  {Siemens}}, \bibinfo {author} {\bibfnamefont {M.~M.}\ \bibnamefont
  {Murnane}}, \bibinfo {author} {\bibfnamefont {H.~C.}\ \bibnamefont
  {Kapteyn}}, \bibinfo {author} {\bibfnamefont {S.}~\bibnamefont {Mathias}},
  \bibinfo {author} {\bibfnamefont {M.}~\bibnamefont {Aeschlimann}}, \bibinfo
  {author} {\bibfnamefont {P.}~\bibnamefont {Grychtol}}, \bibinfo {author}
  {\bibfnamefont {R.}~\bibnamefont {Adam}}, \bibinfo {author} {\bibfnamefont
  {C.~M.}\ \bibnamefont {Schneider}}, \bibinfo {author} {\bibfnamefont {J.~M.}\
  \bibnamefont {Shaw}}, \bibinfo {author} {\bibfnamefont {H.}~\bibnamefont
  {Nembach}}, \ and\ \bibinfo {author} {\bibfnamefont {T.~J.}\ \bibnamefont
  {Silva}},\ }\bibfield  {title} {\enquote {\bibinfo {title} {Ultrafast
  demagnetization dynamics at the $m$ edges of magnetic elements observed using
  a tabletop high-harmonic soft x-ray source},}\ }\href {\doibase
  10.1103/PhysRevLett.103.257402} {\bibfield  {journal} {\bibinfo  {journal}
  {Phys. Rev. Lett.}\ }\textbf {\bibinfo {volume} {103}},\ \bibinfo {pages}
  {257402} (\bibinfo {year} {2009})}\BibitemShut {NoStop}%
\bibitem [{\citenamefont {Vodungbo}\ \emph {et~al.}(2012)\citenamefont
  {Vodungbo}, \citenamefont {Gautier}, \citenamefont {Lambert}, \citenamefont
  {Sardinha}, \citenamefont {Lozano}, \citenamefont {Sebban}, \citenamefont
  {Ducousso}, \citenamefont {Boutu}, \citenamefont {Li}, \citenamefont {Tudu}
  \emph {et~al.}}]{vodungbo2012laser}%
  \BibitemOpen
  \bibfield  {author} {\bibinfo {author} {\bibfnamefont {B.}~\bibnamefont
  {Vodungbo}}, \bibinfo {author} {\bibfnamefont {J.}~\bibnamefont {Gautier}},
  \bibinfo {author} {\bibfnamefont {G.}~\bibnamefont {Lambert}}, \bibinfo
  {author} {\bibfnamefont {A.~B.}\ \bibnamefont {Sardinha}}, \bibinfo {author}
  {\bibfnamefont {M.}~\bibnamefont {Lozano}}, \bibinfo {author} {\bibfnamefont
  {S.}~\bibnamefont {Sebban}}, \bibinfo {author} {\bibfnamefont
  {M.}~\bibnamefont {Ducousso}}, \bibinfo {author} {\bibfnamefont
  {W.}~\bibnamefont {Boutu}}, \bibinfo {author} {\bibfnamefont
  {K.}~\bibnamefont {Li}}, \bibinfo {author} {\bibfnamefont {B.}~\bibnamefont
  {Tudu}},  \emph {et~al.},\ }\bibfield  {title} {\enquote {\bibinfo {title}
  {Laser-induced ultrafast demagnetization in the presence of a nanoscale
  magnetic domain network},}\ }\href@noop {} {\bibfield  {journal} {\bibinfo
  {journal} {Nature communications}\ }\textbf {\bibinfo {volume} {3}},\
  \bibinfo {pages} {999} (\bibinfo {year} {2012})}\BibitemShut {NoStop}%
\bibitem [{\citenamefont {Tengdin}\ \emph {et~al.}(2020)\citenamefont
  {Tengdin}, \citenamefont {Gentry}, \citenamefont {Blonsky}, \citenamefont
  {Zusin}, \citenamefont {Gerrity}, \citenamefont {Hellbrück}, \citenamefont
  {Hofherr}, \citenamefont {Shaw}, \citenamefont {Kvashnin}, \citenamefont
  {Delczeg-Czirjak}, \citenamefont {Arora}, \citenamefont {Nembach},
  \citenamefont {Silva}, \citenamefont {Mathias}, \citenamefont {Aeschlimann},
  \citenamefont {Kapteyn}, \citenamefont {Thonig}, \citenamefont {Koumpouras},
  \citenamefont {Eriksson},\ and\ \citenamefont {Murnane}}]{tengdin2020direct}%
  \BibitemOpen
  \bibfield  {author} {\bibinfo {author} {\bibfnamefont {P.}~\bibnamefont
  {Tengdin}}, \bibinfo {author} {\bibfnamefont {C.}~\bibnamefont {Gentry}},
  \bibinfo {author} {\bibfnamefont {A.}~\bibnamefont {Blonsky}}, \bibinfo
  {author} {\bibfnamefont {D.}~\bibnamefont {Zusin}}, \bibinfo {author}
  {\bibfnamefont {M.}~\bibnamefont {Gerrity}}, \bibinfo {author} {\bibfnamefont
  {L.}~\bibnamefont {Hellbrück}}, \bibinfo {author} {\bibfnamefont
  {M.}~\bibnamefont {Hofherr}}, \bibinfo {author} {\bibfnamefont
  {J.}~\bibnamefont {Shaw}}, \bibinfo {author} {\bibfnamefont {Y.}~\bibnamefont
  {Kvashnin}}, \bibinfo {author} {\bibfnamefont {E.~K.}\ \bibnamefont
  {Delczeg-Czirjak}}, \bibinfo {author} {\bibfnamefont {M.}~\bibnamefont
  {Arora}}, \bibinfo {author} {\bibfnamefont {H.}~\bibnamefont {Nembach}},
  \bibinfo {author} {\bibfnamefont {T.~J.}\ \bibnamefont {Silva}}, \bibinfo
  {author} {\bibfnamefont {S.}~\bibnamefont {Mathias}}, \bibinfo {author}
  {\bibfnamefont {M.}~\bibnamefont {Aeschlimann}}, \bibinfo {author}
  {\bibfnamefont {H.~C.}\ \bibnamefont {Kapteyn}}, \bibinfo {author}
  {\bibfnamefont {D.}~\bibnamefont {Thonig}}, \bibinfo {author} {\bibfnamefont
  {K.}~\bibnamefont {Koumpouras}}, \bibinfo {author} {\bibfnamefont
  {O.}~\bibnamefont {Eriksson}}, \ and\ \bibinfo {author} {\bibfnamefont
  {M.~M.}\ \bibnamefont {Murnane}},\ }\bibfield  {title} {\enquote {\bibinfo
  {title} {Direct light–induced spin transfer between different elements in a
  spintronic heusler material via femtosecond laser excitation},}\ }\href
  {\doibase 10.1126/sciadv.aaz1100} {\bibfield  {journal} {\bibinfo  {journal}
  {Science Advances}\ }\textbf {\bibinfo {volume} {6}},\ \bibinfo {pages}
  {eaaz1100} (\bibinfo {year} {2020})},\ \Eprint
  {http://arxiv.org/abs/https://www.science.org/doi/pdf/10.1126/sciadv.aaz1100}
  {https://www.science.org/doi/pdf/10.1126/sciadv.aaz1100} \BibitemShut
  {NoStop}%
\bibitem [{\citenamefont {De~Groot}\ and\ \citenamefont
  {Kotani}(2008)}]{de2008core}%
  \BibitemOpen
  \bibfield  {author} {\bibinfo {author} {\bibfnamefont {F.}~\bibnamefont
  {De~Groot}}\ and\ \bibinfo {author} {\bibfnamefont {A.}~\bibnamefont
  {Kotani}},\ }\href@noop {} {\emph {\bibinfo {title} {Core level spectroscopy
  of solids}}}\ (\bibinfo  {publisher} {CRC press},\ \bibinfo {year}
  {2008})\BibitemShut {NoStop}%
\bibitem [{\citenamefont {Henke}\ \emph {et~al.}(1993)\citenamefont {Henke},
  \citenamefont {Gullikson},\ and\ \citenamefont {Davis}}]{henke1993x}%
  \BibitemOpen
  \bibfield  {author} {\bibinfo {author} {\bibfnamefont {B.}~\bibnamefont
  {Henke}}, \bibinfo {author} {\bibfnamefont {E.}~\bibnamefont {Gullikson}}, \
  and\ \bibinfo {author} {\bibfnamefont {J.}~\bibnamefont {Davis}},\ }\bibfield
   {title} {\enquote {\bibinfo {title} {X-ray interactions: Photoabsorption,
  scattering, transmission, and reflection at e = 50-30,000 ev, z = 1-92},}\
  }\href {\doibase https://doi.org/10.1006/adnd.1993.1013} {\bibfield
  {journal} {\bibinfo  {journal} {Atomic Data and Nuclear Data Tables}\
  }\textbf {\bibinfo {volume} {54}},\ \bibinfo {pages} {181--342} (\bibinfo
  {year} {1993})}\BibitemShut {NoStop}%
\bibitem [{\citenamefont {Hennes}\ \emph {et~al.}(2021)\citenamefont {Hennes},
  \citenamefont {Rösner}, \citenamefont {Chardonnet}, \citenamefont
  {Chiuzbaian}, \citenamefont {Delaunay}, \citenamefont {Döring},
  \citenamefont {Guzenko}, \citenamefont {Hehn}, \citenamefont {Jarrier},
  \citenamefont {Kleibert}, \citenamefont {Lebugle}, \citenamefont {Lüning},
  \citenamefont {Malinowski}, \citenamefont {Merhe}, \citenamefont {Naumenko},
  \citenamefont {Nikolov}, \citenamefont {Lopez-Quintas}, \citenamefont
  {Pedersoli}, \citenamefont {Savchenko}, \citenamefont {Watts}, \citenamefont
  {Zangrando}, \citenamefont {David}, \citenamefont {Capotondi}, \citenamefont
  {Vodungbo},\ and\ \citenamefont {Jal}}]{hennes2020time}%
  \BibitemOpen
  \bibfield  {author} {\bibinfo {author} {\bibfnamefont {M.}~\bibnamefont
  {Hennes}}, \bibinfo {author} {\bibfnamefont {B.}~\bibnamefont {Rösner}},
  \bibinfo {author} {\bibfnamefont {V.}~\bibnamefont {Chardonnet}}, \bibinfo
  {author} {\bibfnamefont {G.~S.}\ \bibnamefont {Chiuzbaian}}, \bibinfo
  {author} {\bibfnamefont {R.}~\bibnamefont {Delaunay}}, \bibinfo {author}
  {\bibfnamefont {F.}~\bibnamefont {Döring}}, \bibinfo {author} {\bibfnamefont
  {V.~A.}\ \bibnamefont {Guzenko}}, \bibinfo {author} {\bibfnamefont
  {M.}~\bibnamefont {Hehn}}, \bibinfo {author} {\bibfnamefont {R.}~\bibnamefont
  {Jarrier}}, \bibinfo {author} {\bibfnamefont {A.}~\bibnamefont {Kleibert}},
  \bibinfo {author} {\bibfnamefont {M.}~\bibnamefont {Lebugle}}, \bibinfo
  {author} {\bibfnamefont {J.}~\bibnamefont {Lüning}}, \bibinfo {author}
  {\bibfnamefont {G.}~\bibnamefont {Malinowski}}, \bibinfo {author}
  {\bibfnamefont {A.}~\bibnamefont {Merhe}}, \bibinfo {author} {\bibfnamefont
  {D.}~\bibnamefont {Naumenko}}, \bibinfo {author} {\bibfnamefont {I.~P.}\
  \bibnamefont {Nikolov}}, \bibinfo {author} {\bibfnamefont {I.}~\bibnamefont
  {Lopez-Quintas}}, \bibinfo {author} {\bibfnamefont {E.}~\bibnamefont
  {Pedersoli}}, \bibinfo {author} {\bibfnamefont {T.}~\bibnamefont
  {Savchenko}}, \bibinfo {author} {\bibfnamefont {B.}~\bibnamefont {Watts}},
  \bibinfo {author} {\bibfnamefont {M.}~\bibnamefont {Zangrando}}, \bibinfo
  {author} {\bibfnamefont {C.}~\bibnamefont {David}}, \bibinfo {author}
  {\bibfnamefont {F.}~\bibnamefont {Capotondi}}, \bibinfo {author}
  {\bibfnamefont {B.}~\bibnamefont {Vodungbo}}, \ and\ \bibinfo {author}
  {\bibfnamefont {E.}~\bibnamefont {Jal}},\ }\bibfield  {title} {\enquote
  {\bibinfo {title} {Time-resolved xuv absorption spectroscopy and magnetic
  circular dichroism at the ni m2,3-edges},}\ }\href {\doibase
  10.3390/app11010325} {\bibfield  {journal} {\bibinfo  {journal} {Applied
  Sciences}\ }\textbf {\bibinfo {volume} {11}} (\bibinfo {year} {2021}),\
  10.3390/app11010325}\BibitemShut {NoStop}%
\bibitem [{\citenamefont {Chang}\ \emph {et~al.}(2021)\citenamefont {Chang},
  \citenamefont {Guggenmos}, \citenamefont {Cushing}, \citenamefont {Cui},
  \citenamefont {Din}, \citenamefont {Acharya}, \citenamefont {Molesky},
  \citenamefont {Kleineberg}, \citenamefont {Turkowski}, \citenamefont
  {Rahman}, \citenamefont {Neumark},\ and\ \citenamefont
  {Leone}}]{chang2021electron}%
  \BibitemOpen
  \bibfield  {author} {\bibinfo {author} {\bibfnamefont {H.-T.}\ \bibnamefont
  {Chang}}, \bibinfo {author} {\bibfnamefont {A.}~\bibnamefont {Guggenmos}},
  \bibinfo {author} {\bibfnamefont {S.~K.}\ \bibnamefont {Cushing}}, \bibinfo
  {author} {\bibfnamefont {Y.}~\bibnamefont {Cui}}, \bibinfo {author}
  {\bibfnamefont {N.~U.}\ \bibnamefont {Din}}, \bibinfo {author} {\bibfnamefont
  {S.~R.}\ \bibnamefont {Acharya}}, \bibinfo {author} {\bibfnamefont
  {I.~J.~P.}\ \bibnamefont {Molesky}}, \bibinfo {author} {\bibfnamefont
  {U.}~\bibnamefont {Kleineberg}}, \bibinfo {author} {\bibfnamefont
  {V.}~\bibnamefont {Turkowski}}, \bibinfo {author} {\bibfnamefont {T.~S.}\
  \bibnamefont {Rahman}}, \bibinfo {author} {\bibfnamefont {D.~M.}\
  \bibnamefont {Neumark}}, \ and\ \bibinfo {author} {\bibfnamefont {S.~R.}\
  \bibnamefont {Leone}},\ }\bibfield  {title} {\enquote {\bibinfo {title}
  {Electron thermalization and relaxation in laser-heated nickel by
  few-femtosecond core-level transient absorption spectroscopy},}\ }\href
  {\doibase 10.1103/PhysRevB.103.064305} {\bibfield  {journal} {\bibinfo
  {journal} {Phys. Rev. B}\ }\textbf {\bibinfo {volume} {103}},\ \bibinfo
  {pages} {064305} (\bibinfo {year} {2021})}\BibitemShut {NoStop}%
\bibitem [{\citenamefont {Probst}\ \emph {et~al.}(2024)\citenamefont {Probst},
  \citenamefont {M\"oller}, \citenamefont {Schumacher}, \citenamefont {Brede},
  \citenamefont {Dewhurst}, \citenamefont {Reutzel}, \citenamefont {Steil},
  \citenamefont {Sharma}, \citenamefont {Jansen},\ and\ \citenamefont
  {Mathias}}]{PhysRevResearch.6.013107}%
  \BibitemOpen
  \bibfield  {author} {\bibinfo {author} {\bibfnamefont {H.}~\bibnamefont
  {Probst}}, \bibinfo {author} {\bibfnamefont {C.}~\bibnamefont {M\"oller}},
  \bibinfo {author} {\bibfnamefont {M.}~\bibnamefont {Schumacher}}, \bibinfo
  {author} {\bibfnamefont {T.}~\bibnamefont {Brede}}, \bibinfo {author}
  {\bibfnamefont {J.~K.}\ \bibnamefont {Dewhurst}}, \bibinfo {author}
  {\bibfnamefont {M.}~\bibnamefont {Reutzel}}, \bibinfo {author} {\bibfnamefont
  {D.}~\bibnamefont {Steil}}, \bibinfo {author} {\bibfnamefont
  {S.}~\bibnamefont {Sharma}}, \bibinfo {author} {\bibfnamefont {G.~S.~M.}\
  \bibnamefont {Jansen}}, \ and\ \bibinfo {author} {\bibfnamefont
  {S.}~\bibnamefont {Mathias}},\ }\bibfield  {title} {\enquote {\bibinfo
  {title} {Unraveling femtosecond spin and charge dynamics with extreme
  ultraviolet transverse moke spectroscopy},}\ }\href {\doibase
  10.1103/PhysRevResearch.6.013107} {\bibfield  {journal} {\bibinfo  {journal}
  {Phys. Rev. Res.}\ }\textbf {\bibinfo {volume} {6}},\ \bibinfo {pages}
  {013107} (\bibinfo {year} {2024})}\BibitemShut {NoStop}%
\bibitem [{\citenamefont {Hofherr}\ \emph {et~al.}(2020)\citenamefont
  {Hofherr}, \citenamefont {Häuser}, \citenamefont {Dewhurst}, \citenamefont
  {Tengdin}, \citenamefont {Sakshath}, \citenamefont {Nembach}, \citenamefont
  {Weber}, \citenamefont {Shaw}, \citenamefont {Silva}, \citenamefont
  {Kapteyn}, \citenamefont {Cinchetti}, \citenamefont {Rethfeld}, \citenamefont
  {Murnane}, \citenamefont {Steil}, \citenamefont {Stadtmüller}, \citenamefont
  {Sharma}, \citenamefont {Aeschlimann},\ and\ \citenamefont
  {Mathias}}]{hofherr2020ultrafast}%
  \BibitemOpen
  \bibfield  {author} {\bibinfo {author} {\bibfnamefont {M.}~\bibnamefont
  {Hofherr}}, \bibinfo {author} {\bibfnamefont {S.}~\bibnamefont {Häuser}},
  \bibinfo {author} {\bibfnamefont {J.~K.}\ \bibnamefont {Dewhurst}}, \bibinfo
  {author} {\bibfnamefont {P.}~\bibnamefont {Tengdin}}, \bibinfo {author}
  {\bibfnamefont {S.}~\bibnamefont {Sakshath}}, \bibinfo {author}
  {\bibfnamefont {H.~T.}\ \bibnamefont {Nembach}}, \bibinfo {author}
  {\bibfnamefont {S.~T.}\ \bibnamefont {Weber}}, \bibinfo {author}
  {\bibfnamefont {J.~M.}\ \bibnamefont {Shaw}}, \bibinfo {author}
  {\bibfnamefont {T.~J.}\ \bibnamefont {Silva}}, \bibinfo {author}
  {\bibfnamefont {H.~C.}\ \bibnamefont {Kapteyn}}, \bibinfo {author}
  {\bibfnamefont {M.}~\bibnamefont {Cinchetti}}, \bibinfo {author}
  {\bibfnamefont {B.}~\bibnamefont {Rethfeld}}, \bibinfo {author}
  {\bibfnamefont {M.~M.}\ \bibnamefont {Murnane}}, \bibinfo {author}
  {\bibfnamefont {D.}~\bibnamefont {Steil}}, \bibinfo {author} {\bibfnamefont
  {B.}~\bibnamefont {Stadtmüller}}, \bibinfo {author} {\bibfnamefont
  {S.}~\bibnamefont {Sharma}}, \bibinfo {author} {\bibfnamefont
  {M.}~\bibnamefont {Aeschlimann}}, \ and\ \bibinfo {author} {\bibfnamefont
  {S.}~\bibnamefont {Mathias}},\ }\bibfield  {title} {\enquote {\bibinfo
  {title} {Ultrafast optically induced spin transfer in ferromagnetic
  alloys},}\ }\href {\doibase 10.1126/sciadv.aay8717} {\bibfield  {journal}
  {\bibinfo  {journal} {Science Advances}\ }\textbf {\bibinfo {volume} {6}},\
  \bibinfo {pages} {eaay8717} (\bibinfo {year} {2020})},\ \Eprint
  {http://arxiv.org/abs/https://www.science.org/doi/pdf/10.1126/sciadv.aay8717}
  {https://www.science.org/doi/pdf/10.1126/sciadv.aay8717} \BibitemShut
  {NoStop}%
\bibitem [{\citenamefont {M{\"o}ller}\ \emph {et~al.}(2024)\citenamefont
  {M{\"o}ller}, \citenamefont {Probst}, \citenamefont {Jansen}, \citenamefont
  {Schumacher}, \citenamefont {Brede}, \citenamefont {Dewhurst}, \citenamefont
  {Reutzel}, \citenamefont {Steil}, \citenamefont {Sharma},\ and\ \citenamefont
  {Mathias}}]{moller2024verification}%
  \BibitemOpen
  \bibfield  {author} {\bibinfo {author} {\bibfnamefont {C.}~\bibnamefont
  {M{\"o}ller}}, \bibinfo {author} {\bibfnamefont {H.}~\bibnamefont {Probst}},
  \bibinfo {author} {\bibfnamefont {G.~M.}\ \bibnamefont {Jansen}}, \bibinfo
  {author} {\bibfnamefont {M.}~\bibnamefont {Schumacher}}, \bibinfo {author}
  {\bibfnamefont {M.}~\bibnamefont {Brede}}, \bibinfo {author} {\bibfnamefont
  {J.~K.}\ \bibnamefont {Dewhurst}}, \bibinfo {author} {\bibfnamefont
  {M.}~\bibnamefont {Reutzel}}, \bibinfo {author} {\bibfnamefont
  {D.}~\bibnamefont {Steil}}, \bibinfo {author} {\bibfnamefont
  {S.}~\bibnamefont {Sharma}}, \ and\ \bibinfo {author} {\bibfnamefont
  {S.}~\bibnamefont {Mathias}},\ }\bibfield  {title} {\enquote {\bibinfo
  {title} {Verification of ultrafast spin transfer effects in iron-nickel
  alloys},}\ }\href@noop {} {\bibfield  {journal} {\bibinfo  {journal}
  {Communications Physics}\ }\textbf {\bibinfo {volume} {7}},\ \bibinfo {pages}
  {74} (\bibinfo {year} {2024})}\BibitemShut {NoStop}%
\bibitem [{\citenamefont {Oppeneer}(2001)}]{oppeneer2001magneto}%
  \BibitemOpen
  \bibfield  {author} {\bibinfo {author} {\bibfnamefont {P.}~\bibnamefont
  {Oppeneer}},\ }\bibfield  {title} {\enquote {\bibinfo {title}
  {Magneto-optical kerr spectra},}\ }\href@noop {} {\bibfield  {journal}
  {\bibinfo  {journal} {Handbook of Magnetic Materials}\ }\textbf {\bibinfo
  {volume} {13}},\ \bibinfo {pages} {229--422} (\bibinfo {year}
  {2001})}\BibitemShut {NoStop}%
\bibitem [{\citenamefont {Richter}\ \emph {et~al.}(2024)\citenamefont
  {Richter}, \citenamefont {Jana}, \citenamefont {Hennecke}, \citenamefont
  {Schick}, \citenamefont {von Korff~Schmising},\ and\ \citenamefont
  {Eisebitt}}]{richter2024relationship}%
  \BibitemOpen
  \bibfield  {author} {\bibinfo {author} {\bibfnamefont {J.}~\bibnamefont
  {Richter}}, \bibinfo {author} {\bibfnamefont {S.}~\bibnamefont {Jana}},
  \bibinfo {author} {\bibfnamefont {M.}~\bibnamefont {Hennecke}}, \bibinfo
  {author} {\bibfnamefont {D.}~\bibnamefont {Schick}}, \bibinfo {author}
  {\bibfnamefont {C.}~\bibnamefont {von Korff~Schmising}}, \ and\ \bibinfo
  {author} {\bibfnamefont {S.}~\bibnamefont {Eisebitt}},\ }\bibfield  {title}
  {\enquote {\bibinfo {title} {Relationship between magnetic asymmetry and
  magnetization in ultrafast transverse magneto-optical kerr effect
  spectroscopy in the extreme ultraviolet spectral range},}\ }\href {\doibase
  10.1103/PhysRevB.109.184440} {\bibfield  {journal} {\bibinfo  {journal}
  {Phys. Rev. B}\ }\textbf {\bibinfo {volume} {109}},\ \bibinfo {pages}
  {184440} (\bibinfo {year} {2024})}\BibitemShut {NoStop}%
\bibitem [{elk()}]{elk}%
  \BibitemOpen
  \href@noop {} {\enquote {\bibinfo {title} {{The Elk Code}},}\ }\bibinfo
  {howpublished} {\url{http://elk.sourceforge.net/}}\BibitemShut {NoStop}%
\bibitem [{\citenamefont {Edmonds}(1996)}]{edmonds1996angular}%
  \BibitemOpen
  \bibfield  {author} {\bibinfo {author} {\bibfnamefont {A.~R.}\ \bibnamefont
  {Edmonds}},\ }\href@noop {} {\emph {\bibinfo {title} {Angular momentum in
  quantum mechanics}}},\ Vol.~\bibinfo {volume} {4}\ (\bibinfo  {publisher}
  {Princeton university press},\ \bibinfo {year} {1996})\BibitemShut {NoStop}%
\bibitem [{\citenamefont {Dewhurst}\ \emph {et~al.}(2018)\citenamefont
  {Dewhurst}, \citenamefont {Elliott}, \citenamefont {Shallcross},
  \citenamefont {Gross},\ and\ \citenamefont {Sharma}}]{dewhurst2018laser}%
  \BibitemOpen
  \bibfield  {author} {\bibinfo {author} {\bibfnamefont {J.~K.}\ \bibnamefont
  {Dewhurst}}, \bibinfo {author} {\bibfnamefont {P.}~\bibnamefont {Elliott}},
  \bibinfo {author} {\bibfnamefont {S.}~\bibnamefont {Shallcross}}, \bibinfo
  {author} {\bibfnamefont {E.~K.}\ \bibnamefont {Gross}}, \ and\ \bibinfo
  {author} {\bibfnamefont {S.}~\bibnamefont {Sharma}},\ }\bibfield  {title}
  {\enquote {\bibinfo {title} {Laser-induced intersite spin transfer},}\
  }\href@noop {} {\bibfield  {journal} {\bibinfo  {journal} {Nano letters}\
  }\textbf {\bibinfo {volume} {18}},\ \bibinfo {pages} {1842--1848} (\bibinfo
  {year} {2018})}\BibitemShut {NoStop}%
\end{thebibliography}%




\begin{thebibliography}{10}%
\makeatletter
\providecommand \@ifxundefined [1]{%
 \@ifx{#1\undefined}
}%
\providecommand \@ifnum [1]{%
 \ifnum #1\expandafter \@firstoftwo
 \else \expandafter \@secondoftwo
 \fi
}%
\providecommand \@ifx [1]{%
 \ifx #1\expandafter \@firstoftwo
 \else \expandafter \@secondoftwo
 \fi
}%
\providecommand \natexlab [1]{#1}%
\providecommand \enquote  [1]{``#1''}%
\providecommand \bibnamefont  [1]{#1}%
\providecommand \bibfnamefont [1]{#1}%
\providecommand \citenamefont [1]{#1}%
\providecommand \href@noop [0]{\@secondoftwo}%
\providecommand \href [0]{\begingroup \@sanitize@url \@href}%
\providecommand \@href[1]{\@@startlink{#1}\@@href}%
\providecommand \@@href[1]{\endgroup#1\@@endlink}%
\providecommand \@sanitize@url [0]{\catcode `\\12\catcode `\$12\catcode
  `\&12\catcode `\#12\catcode `\^12\catcode `\_12\catcode `\%12\relax}%
\providecommand \@@startlink[1]{}%
\providecommand \@@endlink[0]{}%
\providecommand \url  [0]{\begingroup\@sanitize@url \@url }%
\providecommand \@url [1]{\endgroup\@href {#1}{\urlprefix }}%
\providecommand \urlprefix  [0]{URL }%
\providecommand \Eprint [0]{\href }%
\providecommand \doibase [0]{http://dx.doi.org/}%
\providecommand \selectlanguage [0]{\@gobble}%
\providecommand \bibinfo  [0]{\@secondoftwo}%
\providecommand \bibfield  [0]{\@secondoftwo}%
\providecommand \translation [1]{[#1]}%
\providecommand \BibitemOpen [0]{}%
\providecommand \bibitemStop [0]{}%
\providecommand \bibitemNoStop [0]{.\EOS\space}%
\providecommand \EOS [0]{\spacefactor3000\relax}%
\providecommand \BibitemShut  [1]{\csname bibitem#1\endcsname}%
\let\auto@bib@innerbib\@empty
\bibitem [{\citenamefont {Runge}\ and\ \citenamefont
  {Gross}(1984)}]{runge1984density}%
  \BibitemOpen
  \bibfield  {author} {\bibinfo {author} {\bibfnamefont {E.}~\bibnamefont
  {Runge}}\ and\ \bibinfo {author} {\bibfnamefont {E.~K.~U.}\ \bibnamefont
  {Gross}},\ }\href {\doibase 10.1103/PhysRevLett.52.997} {\bibfield  {journal}
  {\bibinfo  {journal} {Phys. Rev. Lett.}\ }\textbf {\bibinfo {volume} {52}},\
  \bibinfo {pages} {997} (\bibinfo {year} {1984})}\BibitemShut {NoStop}%
\bibitem [{\citenamefont {Dewhurst}\ \emph {et~al.}(2020)\citenamefont
  {Dewhurst}, \citenamefont {Willems}, \citenamefont {Elliott}, \citenamefont
  {Li}, \citenamefont {Schmising}, \citenamefont {Str\"uber}, \citenamefont
  {Engel}, \citenamefont {Eisebitt},\ and\ \citenamefont
  {Sharma}}]{smdewhurst2020element}%
  \BibitemOpen
  \bibfield  {author} {\bibinfo {author} {\bibfnamefont {J.~K.}\ \bibnamefont
  {Dewhurst}}, \bibinfo {author} {\bibfnamefont {F.}~\bibnamefont {Willems}},
  \bibinfo {author} {\bibfnamefont {P.}~\bibnamefont {Elliott}}, \bibinfo
  {author} {\bibfnamefont {Q.~Z.}\ \bibnamefont {Li}}, \bibinfo {author}
  {\bibfnamefont {C.~v.~K.}\ \bibnamefont {Schmising}}, \bibinfo {author}
  {\bibfnamefont {C.}~\bibnamefont {Str\"uber}}, \bibinfo {author}
  {\bibfnamefont {D.~W.}\ \bibnamefont {Engel}}, \bibinfo {author}
  {\bibfnamefont {S.}~\bibnamefont {Eisebitt}}, \ and\ \bibinfo {author}
  {\bibfnamefont {S.}~\bibnamefont {Sharma}},\ }\href {\doibase
  10.1103/PhysRevLett.124.077203} {\bibfield  {journal} {\bibinfo  {journal}
  {Phys. Rev. Lett.}\ }\textbf {\bibinfo {volume} {124}},\ \bibinfo {pages}
  {077203} (\bibinfo {year} {2020})}\BibitemShut {NoStop}%
\bibitem [{\citenamefont {Ryan}\ \emph {et~al.}(2023)\citenamefont {Ryan},
  \citenamefont {Johnsen}, \citenamefont {Elhanoty}, \citenamefont {Grafov},
  \citenamefont {Li}, \citenamefont {Delin}, \citenamefont {Markou},
  \citenamefont {Lesne}, \citenamefont {Felser}, \citenamefont {Eriksson},
  \citenamefont {Kapteyn}, \citenamefont {Grånäs},\ and\ \citenamefont
  {Murnane}}]{smryan2023optically}%
  \BibitemOpen
  \bibfield  {author} {\bibinfo {author} {\bibfnamefont {S.~A.}\ \bibnamefont
  {Ryan}}, \bibinfo {author} {\bibfnamefont {P.~C.}\ \bibnamefont {Johnsen}},
  \bibinfo {author} {\bibfnamefont {M.~F.}\ \bibnamefont {Elhanoty}}, \bibinfo
  {author} {\bibfnamefont {A.}~\bibnamefont {Grafov}}, \bibinfo {author}
  {\bibfnamefont {N.}~\bibnamefont {Li}}, \bibinfo {author} {\bibfnamefont
  {A.}~\bibnamefont {Delin}}, \bibinfo {author} {\bibfnamefont
  {A.}~\bibnamefont {Markou}}, \bibinfo {author} {\bibfnamefont
  {E.}~\bibnamefont {Lesne}}, \bibinfo {author} {\bibfnamefont
  {C.}~\bibnamefont {Felser}}, \bibinfo {author} {\bibfnamefont
  {O.}~\bibnamefont {Eriksson}}, \bibinfo {author} {\bibfnamefont {H.~C.}\
  \bibnamefont {Kapteyn}}, \bibinfo {author} {\bibfnamefont {O.}~\bibnamefont
  {Grånäs}}, \ and\ \bibinfo {author} {\bibfnamefont {M.~M.}\ \bibnamefont
  {Murnane}},\ }\href {\doibase 10.1126/sciadv.adi1428} {\bibfield  {journal}
  {\bibinfo  {journal} {Science Advances}\ }\textbf {\bibinfo {volume} {9}},\
  \bibinfo {pages} {eadi1428} (\bibinfo {year} {2023})},\ \Eprint
  {http://arxiv.org/abs/https://www.science.org/doi/pdf/10.1126/sciadv.adi1428}
  {https://www.science.org/doi/pdf/10.1126/sciadv.adi1428} \BibitemShut
  {NoStop}%
\bibitem [{\citenamefont {Lojewski}\ \emph {et~al.}(2023)\citenamefont
  {Lojewski}, \citenamefont {Elhanoty}, \citenamefont {Le~Guyader},
  \citenamefont {Gr{\aa}n{\"a}s}, \citenamefont {Agarwal}, \citenamefont
  {Boeglin}, \citenamefont {Carley}, \citenamefont {Castoldi}, \citenamefont
  {David}, \citenamefont {Deiter} \emph {et~al.}}]{smlojewski2023interplay}%
  \BibitemOpen
  \bibfield  {author} {\bibinfo {author} {\bibfnamefont {T.}~\bibnamefont
  {Lojewski}}, \bibinfo {author} {\bibfnamefont {M.~F.}\ \bibnamefont
  {Elhanoty}}, \bibinfo {author} {\bibfnamefont {L.}~\bibnamefont
  {Le~Guyader}}, \bibinfo {author} {\bibfnamefont {O.}~\bibnamefont
  {Gr{\aa}n{\"a}s}}, \bibinfo {author} {\bibfnamefont {N.}~\bibnamefont
  {Agarwal}}, \bibinfo {author} {\bibfnamefont {C.}~\bibnamefont {Boeglin}},
  \bibinfo {author} {\bibfnamefont {R.}~\bibnamefont {Carley}}, \bibinfo
  {author} {\bibfnamefont {A.}~\bibnamefont {Castoldi}}, \bibinfo {author}
  {\bibfnamefont {C.}~\bibnamefont {David}}, \bibinfo {author} {\bibfnamefont
  {C.}~\bibnamefont {Deiter}},  \emph {et~al.},\ }\href@noop {} {\bibfield
  {journal} {\bibinfo  {journal} {Materials Research Letters}\ }\textbf
  {\bibinfo {volume} {11}},\ \bibinfo {pages} {655} (\bibinfo {year}
  {2023})}\BibitemShut {NoStop}%
\bibitem [{\citenamefont {Petersilka}\ \emph {et~al.}(1996)\citenamefont
  {Petersilka}, \citenamefont {Gossmann},\ and\ \citenamefont
  {Gross}}]{petersilka1996excitation}%
  \BibitemOpen
  \bibfield  {author} {\bibinfo {author} {\bibfnamefont {M.}~\bibnamefont
  {Petersilka}}, \bibinfo {author} {\bibfnamefont {U.}~\bibnamefont
  {Gossmann}}, \ and\ \bibinfo {author} {\bibfnamefont {E.}~\bibnamefont
  {Gross}},\ }\href@noop {} {\bibfield  {journal} {\bibinfo  {journal}
  {Physical review letters}\ }\textbf {\bibinfo {volume} {76}},\ \bibinfo
  {pages} {1212} (\bibinfo {year} {1996})}\BibitemShut {NoStop}%
\bibitem [{sme()}]{smelk}%
  \BibitemOpen
  \href@noop {} {\enquote {\bibinfo {title} {{The Elk Code}},}\ }\bibinfo
  {howpublished} {\url{http://elk.sourceforge.net/}}\BibitemShut {NoStop}%
\bibitem [{\citenamefont {Oppeneer}(2001)}]{smoppeneer2001magneto}%
  \BibitemOpen
  \bibfield  {author} {\bibinfo {author} {\bibfnamefont {P.}~\bibnamefont
  {Oppeneer}},\ }\href@noop {} {\bibfield  {journal} {\bibinfo  {journal}
  {Handbook of Magnetic Materials}\ }\textbf {\bibinfo {volume} {13}},\
  \bibinfo {pages} {229} (\bibinfo {year} {2001})}\BibitemShut {NoStop}%
\bibitem [{\citenamefont {Condon}\ and\ \citenamefont
  {Shortley}(1935)}]{condon1935theory}%
  \BibitemOpen
  \bibfield  {author} {\bibinfo {author} {\bibfnamefont {E.~U.}\ \bibnamefont
  {Condon}}\ and\ \bibinfo {author} {\bibfnamefont {G.~H.}\ \bibnamefont
  {Shortley}},\ }\href@noop {} {\emph {\bibinfo {title} {The theory of atomic
  spectra}}}\ (\bibinfo  {publisher} {Cambridge University Press},\ \bibinfo
  {year} {1935})\BibitemShut {NoStop}%
\bibitem [{\citenamefont {Bethe}\ and\ \citenamefont
  {Salpeter}(2013)}]{bethe2013quantum}%
  \BibitemOpen
  \bibfield  {author} {\bibinfo {author} {\bibfnamefont {H.~A.}\ \bibnamefont
  {Bethe}}\ and\ \bibinfo {author} {\bibfnamefont {E.~E.}\ \bibnamefont
  {Salpeter}},\ }\href@noop {} {\emph {\bibinfo {title} {Quantum mechanics of
  one-and two-electron atoms}}}\ (\bibinfo  {publisher} {Springer Science \&
  Business Media},\ \bibinfo {year} {2013})\BibitemShut {NoStop}%
\bibitem [{\citenamefont {Bransden}\ and\ \citenamefont
  {Joachain}(2003)}]{bransden2003physics}%
  \BibitemOpen
  \bibfield  {author} {\bibinfo {author} {\bibfnamefont {B.~H.}\ \bibnamefont
  {Bransden}}\ and\ \bibinfo {author} {\bibfnamefont {C.~J.}\ \bibnamefont
  {Joachain}},\ }\href@noop {} {\emph {\bibinfo {title} {Physics of atoms and
  molecules}}}\ (\bibinfo  {publisher} {Pearson Education India},\ \bibinfo
  {year} {2003})\BibitemShut {NoStop}%
\end{thebibliography}
\bibliographystyle{apsrev4-1}

\clearpage
\section*{References for Supplementary Materials}

 \bibliographystyle{apsrev4-1}

\end{document}